\documentclass{article}

\makeatletter
\newcommand{\dontusepackage}[2][]{%
  \@namedef{ver@#2.sty}{9999/12/31}%
  \@namedef{opt@#2.sty}{#1}}
\makeatother
\dontusepackage{algorithmic}
\dontusepackage{algorithm}
\PassOptionsToPackage{table}{xcolor}

\usepackage{microtype}
\usepackage{graphicx}
\usepackage{subfigure}
\usepackage{booktabs} %

\usepackage{hyperref}

\usepackage[accepted]{icml2023}

\usepackage{amsmath}
\usepackage{amssymb}
\usepackage{mathtools}
\usepackage{amsthm}

\usepackage[capitalize,noabbrev]{cleveref}

\usepackage{complexity}
\usepackage{amssymb}
\usepackage{amsmath}
\usepackage{mathtools}
\usepackage{physics}
\usepackage{xspace}
\usepackage{bm}
\usepackage[inline,shortlabels]{enumitem}
\usepackage{amsthm}
\usepackage{aliascnt}
\usepackage{natbib}

\usepackage{booktabs}
\usepackage{multirow}
\usepackage{multicol}

\usepackage{cleveref}

\usepackage[linesnumbered,ruled,lined,noend]{algorithm2e}
\usepackage{etoolbox}

\makeatletter
    \patchcmd\algocf@Vline{\vrule}{\vrule \kern-0.4pt}{}{}
    \patchcmd\algocf@Vsline{\vrule}{\vrule \kern-0.4pt}{}{}
\makeatother
\SetKwComment{Hline}{}{\vspace{-3mm}\textcolor{gray}{\hrule}\vspace{1mm}}
\definecolor{darkgrey}{gray}{0.3}
\definecolor{commentcolor}{gray}{0.5}
\SetKwComment{Comment}{\color{commentcolor}[$\triangleright$\ }{}
\SetCommentSty{}
\SetNlSty{}{\color{darkgrey}}{}
\SetKwProg{Fn}{function}{}{}
\SetKwProg{Subr}{subroutine}{}{}
\crefalias{AlgoLine}{line}%
\crefname{algocf}{Algorithm}{Algorithms}

\usepackage{tikz}
\usepackage{forest}
\usepackage{tikz-qtree}
\usepackage{linegoal}
\usepackage{float}

\usetikzlibrary{positioning,arrows.meta}

\usepackage{autonum} %
\makeatletter
\autonum@generatePatchedReferenceCSL{Cref}
\autonum@generatePatchedReferenceCSL{cref}
\makeatother

\let\cref\Cref

\newcommand{\delimit}[3]{\newcommand{#1}[1]{\left#2##1\right#3}}
\delimit \ceil \lceil \rceil
\delimit \floor \lfloor \rfloor

\let\op\operatorname
\let\eps\varepsilon
\let\mc\mathcal

\renewcommand{\R}{\mathbb R}

\newcommand{\zo}{\{0,1\}}
\renewcommand{\vec}{\bm}
\renewcommand{\va}{{\vec{a}}}
\newcommand{\vx}{{\vec{x}}}
\newcommand{\vy}{{\vec{y}}}
\renewcommand{\vu}{{\vec{u}}}
\newcommand{\mA}{{\mathbf{A}}}
\newcommand{\mU}{{\mathbf{U}}}
\newcommand{\se}[1]{\overset{#1}{\mathbin{\triangleleft}}}

\let\ip\ev
\let\Root\varnothing

\let\co\conv

\newcommand*\circled[1]{\tikz[baseline=(char.base)]{
            \node[shape=circle,draw,inner sep=1pt] (char) {\scriptsize #1};}}

\theoremstyle{plain}
\newtheorem{theorem}{Theorem}[section]
\newtheorem{proposition}[theorem]{Proposition}
\newtheorem{lemma}[theorem]{Lemma}
\newtheorem{corollary}[theorem]{Corollary}
\theoremstyle{definition}
\newtheorem{definition}[theorem]{Definition}

\theoremstyle{remark}

\newenvironment{ienumerate}{\begin{enumerate*}[{\bf 1)}]}{\end{enumerate*}}

\newcommand{\pone}{{\ensuremath{\color{p1color}\blacktriangle}}\xspace}
\newcommand{\ptwo}{{\ensuremath{\color{p2color}\blacktriangledown}}\xspace}
\let\pmax\pone
\let\pmin\ptwo
\definecolor{p1color}{RGB}{31,119,180}
\definecolor{p2color}{RGB}{255,127,14}
\definecolor{p3color}{RGB}{44,160,44}
\definecolor{p4color}{RGB}{214,39,40}

\makeatletter
\def\NAT@spacechar{~}%
\makeatother

\usepackage{siunitx}

  \icmltitlerunning{Team Belief DAG}
\begin{document}

\twocolumn[
    \icmltitle{Team Belief DAG: Generalizing the Sequence Form to Team Games for Fast Computation of Correlated Team Max-Min Equilibria via Regret Minimization}

    \icmlsetsymbol{equal}{*}

    \begin{icmlauthorlist}
        \icmlauthor{Brian Hu Zhang}{cmu}
        \icmlauthor{Gabriele Farina}{meta}
        \icmlauthor{Tuomas Sandholm}{cmu,sr,om,sm}
    \end{icmlauthorlist}

    \icmlaffiliation{cmu}{Carnegie Mellon University, Pittsburgh, PA, USA}
    \icmlaffiliation{meta}{FAIR, Meta AI}
    \icmlaffiliation{sr}{Strategy Robot, Inc., Pittsburgh, PA, USA}
    \icmlaffiliation{om}{Optimized Markets, Inc., Pittsburgh, PA, USA}
    \icmlaffiliation{sm}{Strategic Machine, Pittsburgh, PA, USA}

    \icmlcorrespondingauthor{Brian Hu Zhang}{bhzhang@cs.cmu.edu}
    \icmlcorrespondingauthor{Gabriele Farina}{gfarina@meta.com}
    \icmlcorrespondingauthor{Tuomas Sandholm}{sandholm@cs.cmu.edu}

    \icmlkeywords{Machine Learning, ICML}

    \vskip 0.3in
]

\printAffiliationsAndNotice{} %

\begin{abstract}

    A classic result in the theory of extensive-form games asserts that the set of strategies available to any perfect-recall player is strategically equivalent to a low-dimensional convex polytope, called the \emph{sequence-form polytope}. Online convex optimization tools operating on this polytope are the current state-of-the-art for computing several notions of equilibria in games, and have been crucial in landmark applications of computational game theory.
    However, when optimizing over the \emph{joint} strategy space of a \emph{team} of players, one cannot use the sequence form to obtain a strategically-equivalent convex description of the strategy set of the team.
    In this paper, we provide new complexity results on the computation of optimal strategies for teams, and propose a new representation, coined \emph{team belief DAG (TB-DAG)}, that describes team strategies as a convex set. %
    The TB-DAG enjoys state-of-the-art parameterized complexity bounds, while at the same time enjoying the advantages of efficient regret minimization techniques. We show that TB-DAG can be exponentially smaller and can be computed exponentially faster than all other known representations, and that the converse is never true.  %
    Experimentally, we show that the TB-DAG, when paired with learning techniques, yields state of the art on a wide variety of benchmark team games.

\end{abstract}

\section{Introduction}

In recent years, much research has been concerned with learning strong strategies for players in extensive-form (\emph{i.e.}, tree-form) games. In those settings, a classic result by \citet{Romanovskii62:Reduction} and \citet{Koller94:Fast} asserts that the set of strategies set of each player admits a strategically equivalent and low-dimensional convex description---called the \emph{sequence form}---provided that the agent has perfect recall, that is, that the agent never forgets about past actions or observations. As a result, learning strong strategies for any perfect-recall agent amounts to a convex optimization problem with dimension polynomial in the game tree size, which is typically solved by online learning. Such a template has been used extensively both in the literature and in the applications of computational game theory.

The study of the computational aspects of strategic decision making in adversarial team games is relatively newer. If the team members can privately communicate (for example, a team of poker players colluding at a table secretly revealing to each other their private hands), the team as a whole can be thought of as a single perfect-recall player, and the sequence form can be used. However, when the team members cannot privately communicate during play, the asymmetry in the observations of the different team members makes the sequence form inapplicable. %
This begs the question of how teams should optimize their strategy jointly when communication is impossible.
Such settings are prevalent in the real world, and examples include recreational games like bridge (in which two teams compete adversarially), collusion at a poker table with no means of communicating privately, military situations with restricted communications, various swindling settings, and many other real-world situations.

In general, a polynomially-sized convex description of the strategy set is unlikely to exist even for a team of \textit{two} members, as computing optimal strategies in such team games is known to be NP-hard~\cite{Koller92:Complexity}, but the exact complexity has, to our knowledge, not been established. In this paper, we sharply characterize the complexity of computing optimal team strategies under two common notions, the {\em correlated team max-min equilibrium} (TMECor) and {\em team max-min equilibrium} (TME): they are complete for the complexity classes $\Delta_2^\P$ and $\Sigma_2^\P$ respectively. Here, the result most similar to ours is from \citet{Koller92:Complexity}, who showed that computing a max-min {\em pure} strategy for a team is also $\Sigma_2^\P$-complete. %

Until recently, the best techniques for solving adversarial team games in absence of communication were based on {\em column generation}~\citep{Farina18:ExAnte,Farina21:Connecting,Zhang21:Computing,Zhang22:Optimal}. Those techniques work well in some small and medium-sized games in practice, but generally have no or weak theoretical guarantees. More recently, \citet{Zhang22:Team} developed an algorithm for solving adversarial team games based on a novel {\em tree decomposition} of each player's strategy space, and use it to devise a linear program. They show parameterized complexity bounds that scale with the amount of uncommon information among team members. Simultaneously, \citet{Carminati21:Public} developed an algorithm for converting the game into a strategically equivalent (but exponentially-larger) two-{\em player} game with perfect recall, inspired by prior research in the multi-agent reinforcement learning community~({\em e.g.},~\citealp{Nayyar13:Decentralized,Sokota21:Solving}).

Our main contribution is a representation that has several advantages over the aforementioned papers~\cite{Zhang22:Team,Carminati21:Public}. It can be exponentially smaller and can take exponentially less time to construct than either constructions---and the reverse is never true. It is also conceptually cleaner, especially compared to \citet{Zhang22:Team}: we explicitly give a construction of the team's strategy space, without needing to appeal to the onerous machinery of tree decompositions; furthermore, our more refined construction saves the need for a non-trivial preprocessing step, namely {\em inflation}.
An in-depth comparison between ours and those prior approaches is in \cref{sec:prior research}.

In experiments, we demonstrate that the state-of-the-art variants of counterfactual regret minimization---namely DCFR~\cite{Brown19:Solving}, LCFR~\cite{Brown19:Solving}, or PCFR$^+$~\cite{Farina21:Faster}---applied on top of our TB-DAG outperforms the prior state of the art on almost every game tested by a large margin.

\section{Preliminaries}\label{se:prelims}

A player in an extensive-form game is faced with a decision problem over a tree $\mc H$ (rooted at some node $\Root \in \mc H$). Each node $h \in \mc H$ is either active (where the player selects an action $a$---that is, an outgoing edge from $h$) or inactive (where someone else, possibly adversarially, selects the action to take). To model imperfect information regarding the actions at inactive nodes, the active nodes are partitioned into {\em information sets}, or {\em infosets} for short. We denote by $\mc I$ the collection of infosets for the player.
Nodes in the same infoset are indistinguishable by the player: the player's action must be the same at all nodes in a given infoset.

    {\bf Notation.} $A_h$ denotes the set of actions available ({\em i.e.,} edges emanating from) at node $h$. We make the standard assumption that the player always knows her legal action set. That is, the action set $A_h$ is the same for every node $h$ in any given infoset $I$, and we denote this common action set $A_I$. The child of node $h$ reached by taking action $a$ be denoted $ha$. Leaves of $\mc H$ are called {\em terminal nodes}, and we denote by $\mc Z$ the set of all terminal nodes. We use $\preceq$ to denote the precedence order induced by the tree: if $h, h' \in \mc S$ are two nodes, $h \preceq h'$ means that there is a directed path from $h$ to $h'$. If $S$ is a set of nodes, then $h \preceq S$ means $h \preceq h'$ for some $h' \in S$, and $S \preceq h$ is defined analogously.

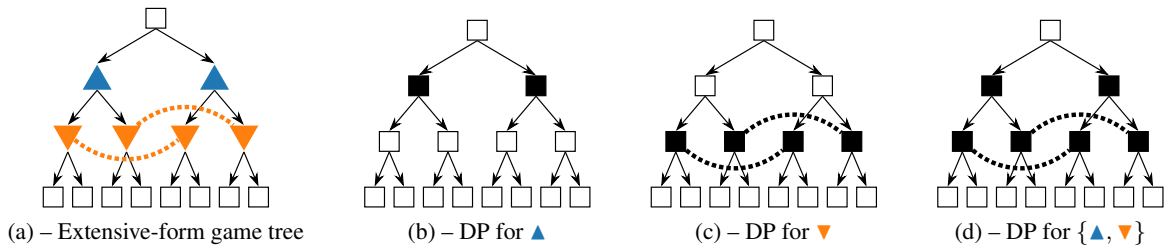
\begin{figure*}[t]
    \centering
\def\sc{1.12}
\tikzset{
 every node/.style={draw, font=\sf},
 every path/.style={-{Stealth[width=1.1mm]}},
 dec/.style={draw, fill=black, text=white},
 pubnode/.style={fill=red, fill opacity=0.2, draw=red},
 infoset/.style={-, densely dotted, line width=1.5pt},
 parent/.style={no edge,tikz={\draw (#1.south) -- (!.north); }},
 circled/.style={draw, circle, minimum size=1.3em, inner sep=0mm},
}
\newcommand{\mymidrulegray}{\arrayrulecolor{gray}\midrule}
\forestset{
    parent/.style={no edge,tikz={\draw (#1.south) -- (!.north); }},
    parent2/.style n args={2}{no edge,tikz={\draw (#1.south) -- (!.north); \draw (#2.south) -- (!.north); }},
    parent3/.style n args={3}{no edge,tikz={\draw (#1.south) -- (!.north); \draw (#2.south) -- (!.north); \draw (#3.south) -- (!.north); }},
    parent4/.style n args={4}{no edge,tikz={\draw (#1.south) -- (!.north); \draw (#2.south) -- (!.north); \draw (#3.south) -- (!.north); \draw (#4.south) -- (!.north); }},
    default preamble={for tree={parent anchor=south, child anchor=north, s sep=1mm,l=0mm}},
        p1/.style={regular
        polygon, regular polygon
        sides=3, inner sep=2pt, fill=p1color, draw=none},
    p2/.style={p1, shape border rotate=180, fill=p2color},
    p3/.style={circle, fill=p3color, draw=none, inner sep=2.8pt}
}
\setlength{\tabcolsep}{4mm}
\begin{tabular}{cccc}
\scalebox{\sc}{\begin{forest}
[,no edge
    [,name=B,p1
        [,p2,name=D
            [] []
        ] 
        [,p2,name=E
            [] []
        ]
    ]
    [,name=C,p1
        [,p2,name=F
            [] []
        ]
        [,p2,name=G
            [] []
        ]
    ]
]
\draw[infoset,bend right=40,p2color] (D) to (F);
\draw[infoset,bend left=40,p2color] (E) to (G);
\end{forest}}
&
\scalebox{\sc}{\begin{forest}
[,no edge
    [,name=B,dec
        [,name=D
            [] []
        ] 
        [,name=E
            [] []
        ]
    ]
    [,name=C,dec
        [,name=F
            [] []
        ]
        [,name=G
            [] []
        ]
    ]
]
\end{forest}}
&
\scalebox{\sc}{\begin{forest}
[,no edge
    [,name=B
        [,dec,name=D
            [] []
        ] 
        [,dec,name=E
            [] []
        ]
    ]
    [,name=C
        [,dec,name=F
            [] []
        ]
        [,dec,name=G
            [] []
        ]
    ]
]
\draw[infoset,bend right=40] (D) to (F);
\draw[infoset,bend left=40] (E) to (G);
\end{forest}}
&
\scalebox{\sc}{\begin{forest}
[,no edge
    [,name=B,dec
        [,dec,name=D
            [] []
        ] 
        [,dec,name=E
            [] []
        ]
    ]
    [,name=C,dec
        [,dec,name=F
            [] []
        ]
        [,dec,name=G
            [] []
        ]
    ]
]
\draw[infoset,bend right=40] (D) to (F);
\draw[infoset,bend left=40] (E) to (G);
\end{forest}}\\
\small (a) -- Extensive-form game tree & \small  (b) -- DP for \pone & \small (c) -- DP for \ptwo & \small (d) -- DP for $\{ \pone, \ptwo \}$
\end{tabular}
    \caption{An example extensive-form game tree with two players (a), and its decision problems for \pone (b), \ptwo (c), and the team consisting of both \pone and \ptwo (d). Dotted lines connect nodes in the same infoset. Note that \pone has perfect information, \ptwo has perfect recall, and the team has neither. In the decision problems, black nodes are active and white inactive.}
    \label{fi:example-game}
\end{figure*}
An example of an extensive-form game tree, and some related decision problems, can be found in \Cref{fi:example-game}. The game tree represents a simple {\em signalling game}, where \pone privately observes the decision of nature and can send a single bit of information to \ptwo. The infosets for \ptwo indicate that \ptwo knows what action \pone played, but {\em not} what action nature played.

A decision problem is {\em timeable} if all paths from the root to a given infoset have the same length. Intuitively,
this means that time is common knowledge. In this paper, we will only consider timeable decision problems. %

    {\bf Realization-Form Representation of Strategies.}
A {\em pure strategy} is a selection of one action at each infoset. %
The {\em realization form} of a pure strategy is the vector $\vx \in \{0,1\}^{\mc Z}$ for which $\vx[z] = 1$ if and only if the strategy prescribes all the actions of that player on the path $\Root \to z$.

A {\em mixed strategy} is a distribution over pure strategies. The realization form $\vx \in [0, 1]^{\mc Z}$ of a mixed strategy is defined as the corresponding convex combination of realization forms of the pure strategies. We will call the set of all realization-form mixed strategies of a player their {\em strategy space}, and denote it $\mc X$. By definition, $\mc X$ is a compact, convex set.

A {\em behavioral strategy} is a mixed strategy in which actions are independently chosen at each infoset. That is, a behavioral strategy is specified by a collection of distributions over actions, one per infoset. We will denote by $\tilde{\mc X}$ the set of realization-form behavioral strategies. The sets $\mc X$ and $\tilde{\mc X}$ do {\em not} coincide in general: indeed, $\tilde{\mc X}$ is often non-convex.

Describing $\mc X$ efficiently (\emph{e.g.}, via a polynomially-sized system of linear constraints) is critical to many modern algorithms for equilibrium finding in games. For example, such a description enables algorithms based on linear programming~\cite{Koller94:Fast} or regret minimization~\cite{Zinkevich07:Regret}, which have been  at the heart of many breakthroughs in computational game theory.

    {\bf Perfect Information.}
We say that a decision problem is {\em perfect-information} if every information set is a singleton. In that case, it is possible to describe the set $\mc X$ efficiently via the following ``probability-flow'' constraints. In the below, $\vec x \in [0, 1]^{\mc H}$, and $\mc X$ is the projection of the resulting polytope onto $[0, 1]^{\mc Z}$. 
\begin{align}
    \left\{~~\begin{aligned}\label{eq:pi-constraints}
        \vx[\Root] & = 1                                                                \\
        \vx[h]     & = \sum_{a \in A_h} \vx[ha] & \forall\, h \text{ active}              \\
        \vx[h]     & = \vx[ha]                  & \forall\, h \text{ inactive}, a \in A_h \\
        \vx[h]     & \ge 0                      & \forall\, h \in \mc H.
    \end{aligned}\right.\quad
\end{align}

{\bf Perfect Recall.}
The {\em sequence} $\sigma(h)$ of a node $h$ is the ordered list of infosets and actions traversed by the player along the path $\Root \to h$. The decision problem has {\em perfect recall} if, at every infoset $I$, every node in $I$ has the same sequence. Intuitively, this means that the player never forgets any information. Perfect-recall decision problems are particularly well-behaved. First, mixed and behavioral strategies coincide under perfect recall---that is, $\mc X = \tilde{\mc X}$. Second, the set of realization-form strategies $\mc X \subseteq [0, 1]^{\mc Z}$ can be expressed efficiently by converting it to an equivalent perfect-{\em information} decision problem known as the {\em sequence form}, as we now make formal.
\begin{definition}
    Two decision problems are {\em strategically equivalent} if they have the same set of terminal nodes $\mc Z$, and the same strategy space $\mc X$.
\end{definition}
\begin{theorem}[\citealp{Romanovskii62:Reduction,Koller94:Fast}]\label{th:sequence form}
    Every perfect-recall decision problem with $n$ nodes is strategically equivalent to a perfect-information decision problem, called its {\em sequence-form} decision problem, with at most $2n$ nodes.
\end{theorem}
Combined with \eqref{eq:pi-constraints}, \cref{th:sequence form} immediately implies the existence of an efficient description of the strategy space in any perfect-recall imperfect-information decision problem.%

    {\bf Imperfect Recall and Teams.}
Decision problems without perfect recall can be thought of from two different, equivalent perspectives. The first, as the name suggests, is the perspective of a single player who sometimes forgets information. The second is that the single player in fact represents a {\em collection} of perfect-recall players that form a team. In this perspective, the player is a {\em team controller}, and (realization-form) mixed strategies are called {\em correlation plans}, because the players' individual strategies need not be independent from each other. The two perspectives are equivalent: clearly, a decision problem for a team can be viewed as an imperfect-recall decision problem by merely ignoring which infosets belong to which team members; conversely, an imperfect-recall decision problem can be viewed as a decision problem for a team of players, where each player $i$ controls a subset $\mc I_i \subseteq \mc I$ of infosets obeying perfect recall. Which perspective is taken is, for our purposes, rather arbitrary; we have chosen to use the perspective of team games because we believe it to be better motivated in practice. We will therefore call a decision problem a {\em team decision problem} if it may lack perfect recall. In the general case, no $\poly(\abs{\mc Z})$-sized description of the strategy space $\mc X$ of a team decision problem can exist unless P = NP~\cite{Koller92:Complexity}. This is despite the fact that $\mc X$ is a convex subset of $[0, 1]^{\mc Z}$.

    {\bf Adversarial Games.}
A {\em zero-sum} or {\em adversarial game} can be described as a tuple $(\mc H, \mc I, \mc J, u)$ where $(\mc H, \mc I)$ and $(\mc H, \mc J)$ are decision problems for two players with a shared underlying tree $\mc H$, and $\vu \in \R^{\mc Z}$ is a utility function for the first player. If the decision problems are team decision problems, the game is an {\em adversarial team game}. The solution concept of interest in adversarial games can be often obtained by solving an appropriate bilinear saddle-point problem, that is, a problem of the form
\begin{align}\label{eq:tmecor}
    \max_{\vx \in \mc X} \min_{\vy \in \mc Y} \vx^\top \mU \vy %
\end{align}
where $\mc X$ and $ \mc Y$ are the mixed strategy spaces of the two players or teams, and $\mU$ is an appropriate payoff matrix.
\begin{definition}
    A {\em team maxmin equilibrium with correlation}, or \textit{TMECor} for short, is a solution to \eqref{eq:tmecor} when $\mc X$ and $\mc Y$ are team decision problems.
\end{definition}

TMECor is not to be confused with {\em team maxmin equilibrium}, which is a related but distinct solution concept that restricts the teams to their behavioral strategies:
\begin{definition}
    A {\em team maxmin equilibrium}, or {\em TME}~\cite{vonStengel97:Team}, is a solution to the  problem
\begin{align}\label{eq:tme}
    \max_{\tilde{\vx} \in \tilde{\mc X}} \min_{\tilde{\vy} \in \tilde{\mc Y}} \tilde{\vx}^\top \mU \tilde{\vy}  %
\end{align}
\end{definition}
Since $\tilde{\mc X}$ and $\tilde{\mc Y}$ are nonconvex (when the players/teams have imperfect recall) in general, swapping the min and max in \eqref{eq:tme} may change the value of the problem---that is, the minimax theorem fails---something that cannot happen with TMECor. In fact, the computation of TME is substantially different from the computation of TMECor and, as we show in \Cref{sec:complexity body}, it is {\em strictly} harder. 

    {\bf Online Convex Optimization.}
    {\em Online convex optimization}~\cite{Zinkevich03:Online} is a framework for describing repeated interactions of a player with an arbitrary environment. At each timestep, the player selects a {\em strategy} $\vx^t$ from a convex, compact set $\mc X$, and observes a (possibly adversarially chosen) {\em utility vector} $\vu^t \in \R^n$. The goal of a regret-minimizing player (a.k.a. a \emph{regret minimizer}) is to ensure that the {\em regret} after $T$ timesteps,
\begin{align}
    R^T := \max_{\vx \in \mc X} \sum_{t=1}^T \ip{\vu^t, \vx - \vx^t},
\end{align}
grow sublinearly in $T$, no matter the sequence $\vu^t$ chosen by the environment. In this paper, we will be concerned with regret minimizers defined over the set $\mc X \subseteq [0, 1]^{\mc Z}$ of realization-form strategies in various decision problems.

Regret minimization is well understood in perfect-information decision problems: {\em counterfactual regret minimization (CFR)}~\cite{Zinkevich07:Regret} is a well-known framework for constructing regret minimizers over $\mc X$ when the decision problem has perfect information.
\begin{theorem}[\citealt{Zinkevich07:Regret}]\label{th:cfr}
    For any perfect-information decision problem, there exists a regret minimizer over the set of realization-form mixed strategies that achieves regret $O(\abs{\mc Z} \sqrt{T})$.
\end{theorem}
This gives a framework for building regret minimizers in many decision problems: construct a strategically-equivalent perfect-information decision problem, and apply CFR. In particular, it immediately implies that regret minimization is possible for perfect-{\em recall} decision problems via the sequence form (\Cref{th:sequence form}).

There is a strong connection between regret minimization and equilibrium computation. Specifically, if the two players play according to regret minimizers on $\mc X$ and $\mc Y$ achieving regrets $R_1$ and $R_2$, respectively, after $T$ timesteps, then a folklore result states that the \emph{average} strategies $\bar{\vx}, \bar{\vy}$  of the two players up until time $T$ satisfy
\begin{align}
    \max_{\vx\in \mc X} \vx^\top \mU \bar{\vy} - \min_{\vy\in \mc Y} \bar{\vx}^\top \mU \vy \le \frac{R_1 + R_2}{T}
    \label{eq:bspp}
\end{align}
implying convergence in the limit to the set of solutions to~\eqref{eq:bspp}.
Equilibrium finding via regret minimization is a key module in game solving, and algorithms that use regret minimization are the practical state of the art in adversarial games~\cite{Brown19:Solving,Farina21:Faster}.

\section{New Complexity Results}\label{sec:complexity body}

In this section, we provide new complexity results regarding finding TMECor and TME values, showing that the computation of the two solution concepts is fundamentally different, especially in team-vs-team games. 

Specifically, we show that for team-vs-team games, computing the TMECor value and computing the TME value are complete for complexity classes $\Delta_2^\P$ and $\Sigma_2^\P$, respectively. These are two complexity classes that form a part of the {\em polynomial hierarchy}. Informally, $\Delta_2^\P$ is the set of decision problems that can be solved in polynomial time given a SAT oracle, and $\Sigma_2^\P$ is the set of decision problems that can be solved in {\em nondeterministic} polynomial time given a SAT oracle. Thus, $\P \subseteq \NP \subseteq \Delta_2^\P \subseteq \Sigma_2^\P$, and all these inclusions are conjectured to be strict. For an overview of the polynomial hierarchy, we refer the reader to Chapter 5 of \citet{Arora09:Computational}.

Interestingly, this distinction does not apply in team-vs-(perfect-recall)-player settings, where both decision problems are \NP-complete.
\begin{center}
    \setlength\tabcolsep{1.1mm}
    \scalebox{.85}{\begin{tabular}{c|>{\centering\arraybackslash}m{3.5cm}|>{\centering\arraybackslash}m{3.5cm}}
                   & Team vs Player                          & Team vs Team                                                                                                                          \\ \toprule
            TMECor & \NP-complete\vskip.1mm \citep{Koller92:Complexity} & {\color{red}$\Delta_2^\P$-complete}\vskip.1mm (This paper,\vskip.1mm Theorems~\ref{th:tmecor-delta2}~and~\ref{th:tmecor-delta2-hard}) \\\midrule
            TME    & \NP-complete\vskip.1mm \citep{Koller92:Complexity} & {\color{red}$\Sigma_2^\P$-complete}\vskip.1mm (This paper,\vskip.1mm Theorems \ref{th:tme-sigma2} and~\ref{th:tme-sigma2-hard})       \\ \bottomrule
        \end{tabular}}
\end{center}

The results above show that, unless the polynomial hierarchy collapses, solving {\em team-vs-player} games, for either concept, is strictly easier than solving {\em team-vs-team} games, and in the team-vs-team setting, TMECor is strictly easier than TME. We believe that this discrepancy suggests that the study of TMECor should use separate techniques from that of TME, especially in the {\em team-vs-team} setting.

\section{Public Observations and the TB-DAG}

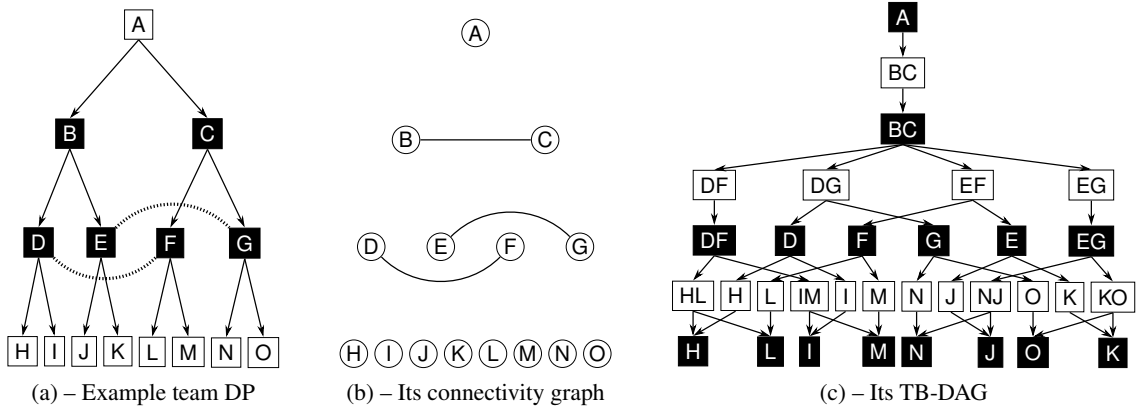
\begin{figure*}[t]
    \centering
    \tikzset{
 every node/.style={draw, font=\sf},
 every path/.style={semithick,-{Stealth[width=1.157mm]}},
 dec/.style={draw, fill=black, text=white},
 pubnode/.style={fill=red, fill opacity=0.2, draw=red},
 infoset/.style={-, densely dotted, line width=1.5pt},
 parent/.style={no edge,tikz={\draw (#1.south) -- (!.north); }},
 circled/.style={draw, circle, minimum size=1.3em, inner sep=0mm},
}
\newcommand{\mymidrulegray}{\arrayrulecolor{gray}\midrule}
\forestset{
    parent/.style={no edge,tikz={\draw (#1.south) -- (!.north); }},
    parent2/.style n args={2}{no edge,tikz={\draw (#1.south) -- (!.north); \draw (#2.south) -- (!.north); }},
    parent3/.style n args={3}{no edge,tikz={\draw (#1.south) -- (!.north); \draw (#2.south) -- (!.north); \draw (#3.south) -- (!.north); }},
    parent4/.style n args={4}{no edge,tikz={\draw (#1.south) -- (!.north); \draw (#2.south) -- (!.north); \draw (#3.south) -- (!.north); \draw (#4.south) -- (!.north); }},
    default preamble={for tree={parent anchor=south, child anchor=north, s sep=1mm,l=0mm}},
}
\setlength{\tabcolsep}{4mm}
\def\Xscale{0.8}
\begin{tabular}{@{}ccc@{}}
\scalebox{\Xscale}{\begin{forest}
for tree={l sep=1.3cm}
[A,no edge
    [B,name=B,dec
        [D,dec,name=D
            [H] [I]
        ] 
        [E,dec,name=E
            [J] [K]
        ]
    ]
    [C,name=C,dec
        [F,dec,name=F
            [L] [M]
        ]
        [G,dec,name=G
            [N] [O]
        ]
    ]
]
\draw[infoset,bend right=45] (D) to (F);
\draw[infoset,bend left=45] (E) to (G);
\end{forest}}
&
\scalebox{\Xscale}{\begin{forest}
for tree={l sep=1.3cm}
[A,for tree={no edge, circled}
    [B,name=B
        [D,name=D
            [H] [I]
        ] 
        [E,name=E
            [J] [K]
        ]
    ]
    [C,name=C
        [F,name=F
            [L] [M]
        ]
        [G,name=G
            [N] [O]
        ]
    ]
]
\draw[-] (B) to (C);
\draw[-,bend right=45] (D) to (F);
\draw[-,bend left=45] (E) to (G);
\end{forest}}
&
\scalebox{\Xscale}{\begin{forest}
[A,dec,no edge
[BC
    [BC,dec
        [DF
            [DF,dec,name=DF
                [HL,parent=DF,name=HL
                    [H,parent2={H2}{HL},dec]
                ]
                [H,parent=D2,name=H2]
            ]
        ]
        [DG,name=DG
            [D,dec,name=D2
                [L,parent=F2,name=L2
                    [L,parent2={L2}{HL},dec]
                ]
                [IM,parent=DF,name=IM
                    [I,parent2={I2}{IM},dec]
                ]
            ]
            [F,dec,parent=EF,name=F2
                [I,parent=D2,name=I2]
                [M,parent=F2,name=M2
                    [M,parent2={M2}{IM},dec]
                ]
            ]
        ]
        [EF,name=EF
            [G,dec,parent=DG,name=G2
                [N,parent=G2,name=N2
                    [N,parent2={N2}{NJ},dec]
                ]
                [J,parent=E2,name=J2]
            ]
            [E,dec,name=E2
                [NJ,parent=EG,name=NJ
                    [J,parent2={J2}{NJ},dec]
                ]
                [O,parent=G2,name=O2
                    [O,parent2={O2}{KO},dec]
                ]
            ]
        ]
        [EG
            [EG,dec,name=EG
                [K,parent=E2,name=K2]
                [KO,parent=EG,name=KO
                    [K,parent2={K2}{KO},dec]
                ]
            ]
        ]
    ]
]
]
\end{forest}}\\
\small (a) -- Example team DP &\small  (b) -- Its connectivity graph &\small  (c) -- Its TB-DAG
\end{tabular}
    \caption{(a) Team decision problem from \Cref{fi:example-game} (nodes named for ease of reference), its connectivity graph (b), and its TB-DAG (c). We remark that per our definition of the construction procedure of TB-DAG, the root is always a decision node; when the root of the original problem is an observation node, this creates a trivial layer in the decision tree. (Of course, it does not affect the complexity guarantees, and in fact the layer might be removed as a postprocessing step---we do this in the experiments; see also \Cref{se:optimizations}, point 2).}
    \label{fi:example}
\end{figure*}

Our main technical contribution is a generalization of \cref{th:sequence form} to team decision problems. Specifically, we develop an algorithm that, like the sequence form, converts a team decision problem into a strategically-equivalent perfect-information decision problem, and use it to develop regret minimization algorithms for solving team games.

Let $\mc T$ be a team decision problem. We will first define the {\em connectivity graph} $G$, which encodes what information is not public to the team.
\begin{definition}
    The {\em connectivity graph} $G$ is the graph whose nodes are the nodes $\mc H$ of $\mc T$, and whose edges connect any two nodes $h, h'$ in the same layer of the tree such that there is an infoset $I$ for the team with $h \preceq I$ and $h' \preceq I$.
\end{definition}
The {\em team belief DAG (TB-DAG)} of a team decision problem $\mc T = (\mc H, \mc I)$ is a perfect-information decision problem $\mc D$
whose active nodes are labeled with specific {\em subsets} of $\mc H$ that intuitively enumerate the possible joint states of all team members, and whose actions intuitively represent all legal combinations of actions that team members can take or observe. Formally, we define $\mc D$ recursively as follows:
\begin{enumerate}[left=1mm,itemsep=\parsep]
    \item The root of $\mc D$ is the active node whose label is the singleton set $\{\Root\}$, containing only the root node of $\mc T$.
    \item The actions available at an active node $s$ of $\mc D$ are defined as follows.
          Let $B \subseteq \mc H$ be the label of the active node.
          If $B$ is a singleton containing a terminal node $z \in \mc Z$, then $s$ is also terminal. Otherwise, let $I_1, \dots, I_m$ be all the infosets in $\mc T$ with nonempty intersection with $B$, and let $J \subseteq \mc H$ be the set of inactive nodes of $\mc T$ in set $B$. The action set at $s$ is the set of {\em prescriptions} $\va \in \bigtimes_{i \in [m]} A_{I_i}$, consisting of one action $a_i$ in each infoset $I_i$. The child reached from $s$ by selecting the prescription $\va$ is an inactive node whose label, which we denote $B \va$, is the set of children of $B$ consistent with $\va$. In symbols:
          $
              B\va := \{ ha_i : h \in I_i \cap B\} \cup \{ h\tilde a : h \in J, \tilde a \in A_h \}.
          $
    \item The actions available at an inactive node $s$ of $\mc D$ are defined as follows. Let $O \subseteq \mc H$ be the label of $s$, and $P_1, \dots, P_m \subseteq O$ be the connected components of the subgraph $G[O]$ of $G$ induced by $O$. We will call $P_1,\dots,P_m$ the {\em public observations} at $O$. The children of $O$ are the active nodes with labels $P_i$ for $i \in [m]$.
\end{enumerate}

The above is a full description of the algorithm for building the TB-DAG; for reference, we include pseudocode for the algorithm in the appendix (\Cref{al:team-dag}). For an example, see the TB-DAG of \cref{fi:example}(a) given in \Cref{fi:example}(c).  The TB-DAG is a decision problem defined on a {\em DAG}, not a {\em tree}.
Despite this, it is valid to discuss decision problems on DAGs in an analogous way to the decision problem on a tree. We formalize DAG decision problems in \Cref{se:dp-dag}.

The active nodes of $\mc D$ are called {\em beliefs}. There can be at most one belief and one inactive node with any given label; therefore, we will refer to them by their labels, {\em e.g.,} {\em the belief $B$} or {\em the inactive node $O$}.

The terminal nodes of $\mc D$ are singleton beliefs, each consisting of one terminal node of $\mc T$. We will therefore identify these two sets of terminal nodes with each other in the natural way, and use $\mc Z$ to refer to the common set of terminal nodes of both decision problems. %
The following result is central in our discussion.\footnote{All proofs are in \Cref{se:proofs} unless otherwise stated.}

\begin{theorem}\label{th:dag equivalence}
    $\mc D$ and $\mc T$ are strategically equivalent.
\end{theorem}

By constructing the TB-DAG, we have traded the presence of {\em infosets} and imperfect recall in $\mc T$ for a (possibly) exponentially larger representation (the TB-DAG). However, as we show, the perfect-information nature of the TB-DAG enables use of online convex optimization methods to compute team equilibria, akin to what the sequence form affords in perfect-recall games. Formally, we have the following.

\begin{theorem}\label{co:cfr eq gap}
    A DAG version of the CFR regret minimizer can be defined and run efficiently in the size of the TB-DAG. Specifically, if the TB-DAG has $N$ nodes and $E$ edges, then the regret of CFR after $T$ iterations is  $O(N\sqrt{T})$, and each iteration takes time $O(E)$.
\end{theorem}

We give full pseudocode for the DAG version of CFR in \cref{al:team-dag}.

\begin{algorithm}[!h]
    \caption{Constructing the TB-DAG.}\label{al:team-dag}
    \DontPrintSemicolon
    \Fn{\normalfont{\sc MakeActiveNode}($B \subseteq \mc H$)}{
        \uIf{\normalfont $\mc D$ has active node with belief $B$}{
            \Return it
        }
        $s \gets{}$new active node in $\mc D$\;
        \If{\normalfont $B = \{ z \} $ for $ z \in \mc Z$}{%
            make $s$ a terminal node\;
            \Return $s$\;
        }
        $\{I_1, \dots, I_m\} \gets \{ I \ni h : h \in B, I \in \mc I\}$\;
        $J \gets \{ h \in B : h \text{ is inactive}\}$\;
        \For{each prescription $\va \in \bigtimes_{i \in [m]} A_{I_i}$}{
            $B \va \gets \{ ha_i : h \in I_i \cap B\} \cup \{ h\tilde a : h \in J, \tilde a \in A_h \}$\;
            add edge $s \to \textsc{MakeInactiveNode}(B \va)$\;
        }
        \Return $s$\;
    }
    \Hline{}
    \Fn{\normalfont{\sc MakeInactiveNode}($O \subseteq \mc H$)}{
        $s \gets{}$new inactive node in $\mc D$\;
        \For{each connected component $P$ of $G[O]$}{
            add edge $s \to \textsc{MakeActiveNode}(P)$
        }
        \Return $s$\;
    }
    \Hline{}
    \Fn{\normalfont{\sc MakeTBDAG}(team decision problem $\mc T$)}{
        {\sc MakeActiveNode}($\{\Root\}$)
    }
\end{algorithm}

In \cref{se:dag size} we give guarantees for the size of the TB-DAG as a function of the amount of uncommon team information, adapting a technique by \citet{Zhang22:Team}. Those bounds immediately imply fixed-parameter bounds for CFR, matching those of \citet{Zhang22:Team}.
Finally, in the next sections we contrast our TB-DAG with prior attempts at obtaining a convex description of the strategy set of a team decision problem, showing that the TB-DAG can be exponentially smaller than all prior descriptions, and that the converse never holds.

\section{Closely-Related Research}\label{sec:prior research}

This paper combines, and at the same time extends, two recent advances in the understanding of the computational aspects surrounding team games. \citet{Carminati21:Public} observed that regret minimization methods can be applied to compute TMECor via a team-belief-based representation. \citet{Zhang22:Team} were the first to point out that it is possible to compute team equilibria with complexity clearly parameterized in the amount of uncommon information in each team. In this section, we delve deeper into the connections between our paper and those two prior results, and discuss how our approach improves over both.

Both of those prior papers use public {\em states} instead of public {\em observations}: that is, in our formalism, their representations would assume only that the team observes a public {\em state} $P \ni h$ at every inactive node, not a public observation. We discuss this distinction in depth in \Cref{se:pubobs-vs-state}.%

\citet{Zhang22:Team} use a formulation based on {\em tree decompositions} to construct a constraint system that describes the polytope of correlated strategies of a team. The constraint system essentially describes what we have called the public belief TB-DAG. Beyond the public state/observation distinction discussed above, \citet{Zhang22:Team} do not discuss the hierarchical, DAG structure of the constraint system we study in this paper, and which we exploited to obtain a CFR-based algorithm for TMECor. In other words, the observation that one can combine online optimization methods while retaining the best parametrized complexity results of tree-decomposition-based methods is novel in this paper.

Instead of focusing on the team members' individual decision problems, \citet{Carminati21:Public} use beliefs to define a {\em converted game}, which is a two-player zero-sum game that is strategically equivalent to the original adversarial team game. Critically, their game is an extensive-form game {\em tree}, and therefore the number of nodes in this tree exceeds the number of {\em paths} through our team belief DAG. Therefore, their converted game can be exponentially larger than our team belief DAG (\Cref{se:carminati counterexample} of our paper shows an explicit construction in which this is the case). They represent the strategy space of the teams via a safe imperfect-recall abstraction of the converted two-player (not two team) zero-sum game~\cite{Lanctot12:No}, resulting in a representation of {\em each team's strategy space} that is, again, essentially the public state TB-DAG. However, since their algorithms operate on the converted game, their time complexity depends on the size of the converted game. Therefore, the algorithms in our paper can be exponentially faster even if the strategy space representation has the same size.

\section{Public States and Observations}\label{se:pubobs-vs-state}

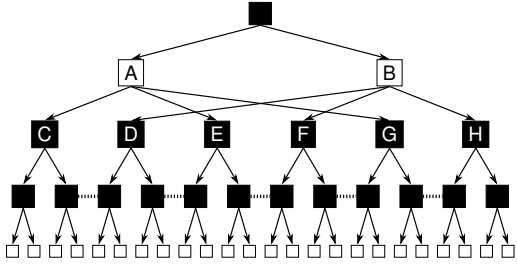
\begin{figure}[t]
\tikzset{
    every path/.style={semithick,-{Stealth[width=1.157mm]}},
    every node/.style={draw=black, fill=white, inner sep=3pt},
    pubnode/.style={fill=red, fill opacity=0.2, draw=red},
    infoset/.style={-, densely dotted, line width=1.7pt},
    circled/.style={draw, circle, minimum size=1.3em, inner sep=0mm},
}
\newcommand{\mymidrulegray}{\arrayrulecolor{gray}\midrule}
\forestset{
    dec/.style={draw, fill=black, text=white, font=\sf, minimum size=2.5ex},
    ds/.style={draw, fill=black, text=white, font=\sf, minimum size=2.5ex},
    obs/.style={minimum size=2.5ex, font=\sf},
    terminal/.style={}, %
    p1/.style={regular
        polygon, regular polygon
        sides=3, inner sep=2pt, fill=p1color, draw=none},
    p2/.style={p1, shape border rotate=180, fill=p2color},
    parent/.style={no edge,tikz={\draw (#1.south) -- (!.north); }},
}
    \centering
\scalebox{.75}{\begin{forest}
for tree={parent anchor=south, child anchor=north, s sep=0.15cm, l sep=0.6cm
}
[,dec
[A,name=A,obs
    [C,dec,parent=A
        [,ds [] []] 
        [,ds,name=1 [] []]
    ]
    [D,dec,parent=B
        [,ds,name=2 [] []] 
        [,ds,name=3 [] []]
    ]
    [E,dec,parent=A
        [,ds,name=4 [] []] 
        [,ds,name=5 [] []]
    ]
]
[B,name=B,obs
    [F,dec,parent=B
        [,ds,name=6 [] []] 
        [,ds,name=7 [] []]
    ]
    [G,dec,parent=A
        [,ds,name=8 [] []] 
        [,ds,name=9 [] []]
    ]
    [H,dec,parent=B
        [,ds,name=10 [] []] 
        [,ds,name=11 [] []]
    ]
]
]
\draw[infoset] (1) to (2);
\draw[infoset] (3) to (4);
\draw[infoset] (5) to (6);
\draw[infoset] (7) to (8);
\draw[infoset] (9) to (10);
\end{forest}}
\vspace{-1mm}
\caption{A team decision problem showing that public-state-based approaches do not subsume inflation.}
    \label{fi:inflation-counterexample}
\vspace{-2mm}
\end{figure}

Both prior approaches use the concept of public {\em states} to construct a strategy space representation. Instead, we propose and use public {\em observations}. In this section, we discuss this difference in depth. Intuitively, the difference is that public observations are {\em localized} to a particular node in the TB-DAG: if a fact is public to the team {\em conditional on the part of the team strategy that has been played to reach this point}, then it is a public observation. On the other hand, public {\em states} only encode information that is {\em unconditionally} public. We now formally define public states.

\begin{definition}\label{de:public state}
    A {\em public state} is a connected component of the connectivity graph $G$.
\end{definition}

We envision an alternative construction of the TB-DAG in which, instead of picking a connected component of $G[O]$ ({\em i.e.,} a public observation), the environment picks a public state $P$ intersecting with $O$ and transitions to $P \cap O$. We will call this version the {\em public state TB-DAG}.

Our first result is that the TB-DAG can never be too much larger than the public state TB-DAG:

\begin{proposition}\label{pr:public state comparison}
    Let $N$ and $N'$ be the number of nodes in the TB-DAG and public state TB-DAG respectively. Then $N \le 2p N'$, where $p$ is the largest size (in number of nodes) of any belief in the public state TB-DAG.
\end{proposition}

Thus, using public observations is never much worse than using public states---and, in practice, it is almost always {\em better}. In the remainder of the section, we will discuss why we strictly prefer using public observations, from both conceptual and theoretical perspectives.

First, using public observations removes the need to \emph{inflate} the information partition of the team before the new representation can be constructed. {\em Complete inflation}~\cite{Kaneko95:Behavior}, which we simply call {\em inflation} for short, is an algorithm that splits an infoset $I$ into two infosets $I = I_1 \sqcup I_2$ if no pure strategy of the team can simultaneously play to a node in $I_1$ and a node in $I_2$, and repeats this process until no more such splits are possible. This preserves strategic equivalence. However, inflation can lead to the break-up of public states, in turn resulting in a reduction in the public state TB-DAG size. 

Indeed, consider the team decision problem in \Cref{fi:inflation-counterexample}. Due to the information sets marked in the last black layer of the game tree, the connectivity graph contains a path C---D---E---...---H. Therefore, \{C, D, ..., H\} form a public state. Also, it is possible for the combinations CEG and DFH to be reached (if the player at the root plays left or right, respectively). Therefore, CEG and DFH are beliefs in the public-state TB-DAG. In the public-observation TB-DAG, consider for example what happens if the left action is played at the root, so that C, E, and G are all reached. Note that there are no edges connecting C, E, and G---the path connecting C to E in the connectivity graph passes through D, which is not reached; therefore, C, E, and G are three different public observations and hence three different beliefs, resulting in an exponentially-smaller TB-DAG. Inflation would remove the nontrivial information sets in the second black layer, which would ultimately have the same effect in this example as using public observations. 

The number $3$ is not special in this construction;  it can be increased arbitrarily by  simply increasing the number of children of {\sf A} and {\sf B}. Therefore, in particular, one can construct a family of games in which the public state TB-DAG (without inflation) has exponential size, while our (public observation) TB-DAG has polynomial size. This is why \citet{Zhang22:Team} and \citet{Carminati21:Public} insist that inflation be done as a preprocessing step before beginning their constructions. The use of public observations, however, removes the need for this step:
\begin{proposition}\label{pr:inflate1}
    Given any team decision problem $\mathcal{T}$, the TB-DAG of $\mathcal{T}$ is the same no matter whether inflation is applied to $\mc T$ before the construction.
\end{proposition}

Although inflation {\em can} be performed efficiently, not requiring it as a preprocessing step simplifies the code and makes for a conceptually cleaner construction.
However, the benefits of public observations go beyond making inflation unnecessary. In fact, even with inflation, there are still cases in which using public observations instead represents an exponential improvement.
\begin{proposition}\label{pr:inflate2}
    There exists a family of team decision problems in which the TB-DAG has polynomial size, but the public state TB-DAG has exponential size, even if inflation is applied as a preprocessing step before building the latter.
\end{proposition}
The construction that proves \Cref{pr:inflate2} is similar to \Cref{fi:inflation-counterexample} but more involved, and is available in \cref{se:proofs}.

\section{Experiments}
\begin{table*}[t]
\newcommand{\cbox}[2]{\fcolorbox{white}[rgb]{#1}{\phantom{28m 13s}\llap{#2}}}
\centering
\newcommand{\mymidrulegray}{\arrayrulecolor{gray}\mymidrule}
\newcommand{\mymidrule}{
\cmidrule(lr){1-5}
\cmidrule(lr){6-7}
\cmidrule(lr){8-9}
\cmidrule(lr){10-10}
\cmidrule(lr){11-15}
\arrayrulecolor{black}}
\newcommand{\unk}{\textcolor{black!30}{---}}
\setlength{\tabcolsep}{1.5mm}
\colorlet{teamcol}{gray}
\newcommand{\teamprint}[1]{\{\textcolor{teamcol}{#1}\}}

\scalebox{.72}{
\begin{tabular}{lrrrrrrrrr|r@{\hskip1mm}rrr@{\hskip1mm}r}
\toprule
      \multirow{2}{*}{\makebox[15mm][l]{{\bf Game} $\{\pmin\}$}} &
      \multirow{2}{*}{\bf Leaves} &
      \llap{\bf Team \pmax} &
      \llap{\bf P. S.} &
      \multirow{2}{*}{$\bm{k}$} &
      \multicolumn{2}{c}{\bf Team \pmax's DAG} &
      \multicolumn{2}{c}{\bf Team \pmin's DAG} &
      {\bf CCCG22} &
      \multicolumn{2}{c}{\bf This paper (CFR)} &
      \!\!{\bf ZS22 (LP)} &
      \multicolumn{2}{c}{\bf ZFCS22 (CG)}
      \\
      &
      &
      \bf Value &
      \bf Size &
      & %
      Vertices &
      Edges &
      Vertices &
      Edges &
      Game Size &
      $\eps=10^{-3}$ &
      \!\!$\eps=10^{-4}$ &
      $\ddagger\quad~~~$&
      \!\!$\eps=10^{-3}$ &
      \!\!$\eps=10^{-4}$
      \\ \mymidrulegray
    \rlap{$^3$K3 \teamprint{3}} &      78 &    0.000 &   6 &  6 &       487 &        918 &      37 &        36 & 4,108 &                \cbox{0.7843137254901961,0.7843137254901961,1.0}{0.00s} &                 \cbox{0.7843137254901961,0.7843137254901961,1.0}{0.00s} &                 \cbox{0.7843137254901961,0.7843137254901961,1.0}{0.01s} &                \cbox{0.7843137254901961,0.7843137254901961,1.0}{0.00s} &                 \cbox{0.7843137254901961,0.7843137254901961,1.0}{0.00s} \\
    \rlap{$^3$K4 \teamprint{3}} &     312 & $-$0.042 &  12 &  8 &     2,100 &      6,711 &      49 &        48 &  66,349 &                \cbox{0.7843137254901961,0.7843137254901961,1.0}{0.00s} &                 \cbox{0.7843137254901961,0.7843137254901961,1.0}{0.00s} &  \cbox{0.8310649750096116,0.8310649750096116,0.9824682814302191}{0.02s} &                \cbox{0.7843137254901961,0.7843137254901961,1.0}{0.01s} &    \cbox{0.823683198769704,0.823683198769704,0.9852364475201846}{0.02s} \\
    \rlap{$^3$K6 \teamprint{3}} &   1,560 & $-$0.024 &  30 & 12 &    54,255 &    336,944 &      73 &        72 & 7,002,763 &                \cbox{0.7843137254901961,0.7843137254901961,1.0}{0.03s} &  \cbox{0.8876585928489042,0.8876585928489042,0.9612456747404844}{0.12s} &  \cbox{0.9797001153402537,0.8384467512495194,0.8384467512495194}{1.22s} & \cbox{0.8962706651287966,0.8962706651287966,0.9580161476355248}{0.14s} &  \cbox{0.8962706651287966,0.8962706651287966,0.9580161476355248}{0.14s} \\
    \rlap{$^3$K8 \teamprint{3}} &   4,368 & $-$0.019 &  56 & 16 & 1,783,926 & 15,564,765 &      97 &        96 & 488,157,721 & \cbox{0.9598615916955017,0.8913494809688581,0.8913494809688581}{4.73s} &                \cbox{1.0,0.7843137254901961,0.7843137254901961}{32.36s} &                \cbox{1.0,0.7843137254901961,0.7843137254901961}{3m 23s} &                \cbox{0.7843137254901961,0.7843137254901961,1.0}{0.23s} &  \cbox{0.8052287581699347,0.8052287581699347,0.9921568627450981}{0.32s} \\
 \rlap{$^{3}$K12 \teamprint{3}} &  17,160 & $-$0.014 & 132 & 24 &      \unk &       \unk &    \unk &      \unk & \unk &                  \cbox{1.0,0.7843137254901961,0.7843137254901961}{oom} &                   \cbox{1.0,0.7843137254901961,0.7843137254901961}{oom} &                   \cbox{1.0,0.7843137254901961,0.7843137254901961}{oom} &                \cbox{0.7843137254901961,0.7843137254901961,1.0}{0.84s} &                 \cbox{0.7843137254901961,0.7843137254901961,1.0}{1.39s} \\
  \rlap{$^4$K5 \teamprint{3,4}} &   3,960 & $-$0.037 &  20 & 10 &    26,566 &    124,875 &   4,621 &    15,415 & \unk &                \cbox{0.7843137254901961,0.7843137254901961,1.0}{0.03s} &  \cbox{0.8322952710495963,0.8322952710495963,0.9820069204152249}{0.05s} &  \cbox{0.9695501730103806,0.8655132641291811,0.8655132641291811}{0.79s} &                                                                   \unk &                                                                    \unk \\
    \rlap{$^4$K5 \teamprint{4}} &   3,960 & $-$0.030 &  60 & 15 &   998,471 &  4,658,070 &     121 &       120 & 202,660,366 &                \cbox{0.7843137254901961,0.7843137254901961,1.0}{1.59s} &  \cbox{0.8778162245290273,0.8778162245290273,0.9649365628604383}{6.34s} &                \cbox{1.0,0.7843137254901961,0.7843137254901961}{3m 25s} &                                                                   \unk &                                                                    \unk \\
\mymidrulegray
  \rlap{$^3$L133 \teamprint{3}} &   6,477 &    0.215 &   9 &  6 &    23,983 &     49,005 &     685 &       684 & 1,691,158 &                \cbox{0.7843137254901961,0.7843137254901961,1.0}{0.02s} &  \cbox{0.8556708958093041,0.8556708958093041,0.9732410611303345}{0.05s} &  \cbox{0.9686274509803922,0.8679738562091504,0.8679738562091504}{0.50s} &               \cbox{1.0,0.7843137254901961,0.7843137254901961}{24.89s} &                \cbox{1.0,0.7843137254901961,0.7843137254901961}{45.96s} \\
  \rlap{$^3$L143 \teamprint{3}} &  20,856 &    0.107 &  16 &  8 &   139,964 &    417,027 &   1,201 &     1,200 & 61,983,093 &                \cbox{0.7843137254901961,0.7843137254901961,1.0}{0.10s} &  \cbox{0.8876585928489042,0.8876585928489042,0.9612456747404844}{0.48s} &  \cbox{0.9921568627450981,0.8052287581699347,0.8052287581699347}{7.58s} &                \cbox{1.0,0.7843137254901961,0.7843137254901961}{2m 4s} &                 \cbox{1.0,0.7843137254901961,0.7843137254901961}{6m 3s} \\
  \rlap{$^3$L151 \teamprint{3}} &  10,020 & $-$0.019 &  20 & 10 &   150,707 &    496,196 &   1,501 &     1,500 & \unk  &                \cbox{0.7843137254901961,0.7843137254901961,1.0}{0.18s} &  \cbox{0.8519800076893502,0.8519800076893502,0.9746251441753172}{0.50s} &    \cbox{0.9829296424452134,0.829834678969627,0.829834678969627}{9.30s} & \cbox{0.9543252595155709,0.9061130334486736,0.9061130334486736}{3.06s} & \cbox{0.9930795847750865,0.8027681660899654,0.8027681660899654}{13.98s} \\
  \rlap{$^3$L153 \teamprint{3}} &  51,215 &    0.024 &  25 & 10 &   855,397 &  3,486,091 &   1,861 &     1,860 & 1,973,610,366 &                \cbox{0.7843137254901961,0.7843137254901961,1.0}{1.24s} &  \cbox{0.8778162245290273,0.8778162245290273,0.9649365628604383}{4.94s} &                \cbox{1.0,0.7843137254901961,0.7843137254901961}{4m 24s} &               \cbox{1.0,0.7843137254901961,0.7843137254901961}{7m 23s} &               \cbox{1.0,0.7843137254901961,0.7843137254901961}{28m 13s} \\
  \rlap{$^3$L223 \teamprint{3}} &   8,762 &    0.516 &   4 &  4 &    32,750 &     45,913 &   2,437 &     2,436 & 538,111  &                \cbox{0.7843137254901961,0.7843137254901961,1.0}{0.03s} &  \cbox{0.8470588235294118,0.8470588235294118,0.9764705882352941}{0.08s} &  \cbox{0.9319492502883506,0.9319492502883506,0.9446366782006921}{0.27s} &               \cbox{1.0,0.7843137254901961,0.7843137254901961}{13.48s} &                \cbox{1.0,0.7843137254901961,0.7843137254901961}{18.53s} \\
  \rlap{$^3$L523 \teamprint{3}} & 775,148 &    0.953 &   4 &  4 & 2,911,352 &  4,183,685 & 220,705 &   220,704 & 222,239,487 &               \cbox{0.7843137254901961,0.7843137254901961,1.0}{11.26s} & \cbox{0.8384467512495194,0.8384467512495194,0.9797001153402537}{24.86s} & \cbox{0.9478662053056517,0.9233371780084583,0.9233371780084583}{2m 26s} &               \cbox{1.0,0.7843137254901961,0.7843137254901961}{$>$ 6h} &                \cbox{1.0,0.7843137254901961,0.7843137254901961}{$>$ 6h} \\
\rlap{$^4$L133 \teamprint{3,4}} &  80,322 &    0.147 &   9 &  6 &    79,351 &    158,058 &  75,157 &   155,475 & 277,714,570 &                \cbox{0.7843137254901961,0.7843137254901961,1.0}{0.21s} &    \cbox{0.885198000768935,0.885198000768935,0.9621683967704728}{0.92s} &  \cbox{0.9732410611303345,0.8556708958093041,0.8556708958093041}{7.30s} &                                                                   \unk &                                                                    \unk \\
\mymidrulegray
    \rlap{$^3$D3 \teamprint{3}} &  13,797 &    0.284 &   9 &  6 &    91,858 &    215,967 &   1,522 &     1,521 & \unk &                \cbox{0.7843137254901961,0.7843137254901961,1.0}{0.11s} &  \cbox{0.8716647443291041,0.8716647443291041,0.9672433679354094}{0.40s} &  \cbox{0.9570934256055363,0.8987312572087659,0.8987312572087659}{2.10s} &               \cbox{1.0,0.7843137254901961,0.7843137254901961}{11.05s} &                \cbox{1.0,0.7843137254901961,0.7843137254901961}{11.05s} \\
    \rlap{$^3$D4 \teamprint{3}} & 262,080 &    0.284 &  16 &  8 & 4,043,377 & 13,749,608 &  16,381 &    16,380 & \unk &               \cbox{0.7843137254901961,0.7843137254901961,1.0}{22.54s} & \cbox{0.8778162245290273,0.8778162245290273,0.9649365628604383}{1m 28s} & \cbox{0.9621683967704728,0.8851980007689351,0.8851980007689351}{8m 29s} &               \cbox{1.0,0.7843137254901961,0.7843137254901961}{3h 19m} &                \cbox{1.0,0.7843137254901961,0.7843137254901961}{3h 19m} \\
  \rlap{$^4$D3 \teamprint{2,4}} & 331,695 &    0.200 &   9 &  6 &   514,120 &  1,217,310 & 486,442 & 1,155,144 & \unk &                \cbox{0.7843137254901961,0.7843137254901961,1.0}{2.31s} &  \cbox{0.8322952710495963,0.8322952710495963,0.9820069204152249}{4.70s} & \cbox{0.9769319492502884,0.8458285274894272,0.8458285274894272}{1m 32s} &                                                                   \unk &                                                                    \unk \\
\rlap{$^6$D2 \teamprint{2,4,6}} & 262,080 &    0.072 &   8 &  6 &   254,758 &    457,795 & 218,570 &   389,995 & \unk &                \cbox{0.7843137254901961,0.7843137254901961,1.0}{1.72s} &  \cbox{0.8458285274894272,0.8458285274894272,0.9769319492502884}{4.26s} & \cbox{0.9393310265282584,0.9393310265282584,0.9418685121107266}{16.55s} &                                                                   \unk &                                                                    \unk \\
  \rlap{$^6$D2 \teamprint{4,6}} & 262,080 &    0.265 &  16 &  8 &   991,861 &  2,029,546 &  46,236 &    60,717 & \unk &                \cbox{0.7843137254901961,0.7843137254901961,1.0}{3.80s} & \cbox{0.8569011918492887,0.8569011918492887,0.9727797001153402}{11.09s} & \cbox{0.9649365628604383,0.8778162245290273,0.8778162245290273}{1m 35s} &                                                                   \unk &                                                                    \unk \\
    \rlap{$^6$D2 \teamprint{6}} & 262,080 &    0.333 &  32 & 10 & 3,158,364 &  7,395,885 &   5,551 &     5,550 & \unk &               \cbox{0.7843137254901961,0.7843137254901961,1.0}{30.20s} & \cbox{0.8421376393694733,0.8421376393694733,0.9783160322952711}{1m 11s} & \cbox{0.9557093425605536,0.9024221453287197,0.9024221453287197}{8m 53s} &                                                                   \unk &                                                                    \unk \\
\bottomrule
\end{tabular}

}

\caption{Runtime of our CFR-based algorithm (column `\textbf{This paper}') using the team belief DAG form, compared to the prior state-of-the-art algorithms based on linear programming and column generation by \citet{Zhang22:Team} (`\textbf{ZS22}') and \citet{Zhang22:Optimal} (`\textbf{ZFCS22}') respectively, on several standard parametric benchmark games. See \Cref{sec:exp discussion} for a description of the games, and for a detailed description of the meaning of each column. Missing or unknown values are denoted with `\unk'. For each row, the background color of each runtime column is set proportionally to the ratio with the best runtime for the row, according to the logarithmic color scale \raisebox{-3.2mm}{\includegraphics[scale=.65]{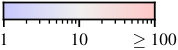}}\!\!\!\!.%
\quad$\ddagger$: Solved to Gurobi's default precision (LP barrier solver does not produce feasible iterates until convergence).}
\label{tab:results}
\end{table*}

We experimentally investigate solving adversarial team games using the team belief DAG. Since all games we experiment on have public actions, we always preprocess with branching factor reduction.

{\bf Algorithms Tested.}\label{sec:exp algos}
We implemented the following state-of-the-art variants of CFR on the TB-DAG: \textit{Predictive CFR$^+$ (PCFR$^+$)}~\citep{Farina21:Faster}, \textit{Discounted CFR (DCFR)}~\citep{Brown19:Solving}, and \textit{Linear CFR (LCFR)}~\citep{Brown19:Solving}. 
    PCFR$^+$ and DCFR use quadratic averaging of iterates, while LCFR uses linear averaging. PCFR$^+$ is a predictive regret minimization algorithm. At each time $t$, we use the previous utility vector for each time as prediction for the next. 
    All implementations are single-threaded.
    
We compare solving an adversarial team games via the team belief DAG against two prior state-of-the-art algorithms:
\begin{ienumerate}
    \item The tree-decomposition-based LP solver proposed by \citet{Zhang22:Team} (henceforth ``\textbf{ZS22}''), which has already been discussed at length in this paper. We used the original implementation of the authors, which internally uses the barrier algorithm implemented by the commercial solver Gurobi. As recommended by the authors, we turned Gurobi's presolver off to avoid numerical instability and increase speed. We allowed Gurobi to use up to four threads.
    \item The single-oracle algorithm of \citet{Farina21:Connecting} (henceforth ``\textbf{ZFCS22}''). ZFCS22 iteratively refines the strategy of each team by solving best-response problems using a tight integer program derived from the theory of extensive-form correlation~\citep{Stengel08:Extensive,Farina21:Faster}. We used the original code by the authors, which was implemented for three-player games in which a team of two players faces an opponent. Like ZS22 and our LP-based solver, ZFCS22 uses the commercial solver Gurobi to solve linear and integer linear programs. We allowed Gurobi to use up to four threads.
\end{ienumerate}

All experiments were run on a 64-core AMD EPYC 7282 processor. Each algorithm was allocated a maximum of 4 threads, 60GBs of RAM, and a time limit of 6 hours. 

{\bf Game Instances.}\label{sec:exp games}
We ran experiments on the following standard, parametric benchmark games:
\begin{ienumerate}
    \item {\bf $^n$K$r$}: $n$-player Kuhn poker with $r$ ranks~\citep{Kuhn50:Simplified}.
    \item {\bf $^n$L$brs$}: $n$-player Leduc poker with a $b$-bet maximum in each betting round, $r$ ranks, and $s$ suits~\citep{Southey05:Bayes}. 
    \item {\bf $^n$D$d$}: $n$-player Liar's Dice with one $d$-sided die for each player~\citep{Lisy15:Online}. 
\end{ienumerate}
These are the same games used by \citet{Zhang22:Team} and \citet{Farina21:Connecting} in their experimental evaluations. We refer the reader to the latter paper for detailed descriptions of the games. The size of each game, measured in terms of number of terminal states (leaves), is reported in the second column of \Cref{tab:results}.

{\bf Discussion of Experimental Results.}\label{sec:exp discussion}
Experimental results are summarized in \Cref{tab:results}. Column `\textbf{Game}` indicates the game, and the set of players on Team~\pmin.  Column `\textbf{P. S. Size}' reports the largest effective size $p$ of any public state. Column `$\bm{k}$' reports the value of $k$ for which both teams are $k$-private. Columns `\textbf{Team $\pmax$'s DAG}' and `\textbf{Team $\pmin$'s DAG}' report the total number of vertices and edges in the team belief DAG for teams $\pmax$ and $\pmin$ respectively. Column `\textbf{Team $\pmax$ value}' reports the utility that team $\pmax$ can expect to gain at equilibrium. Column `\textbf{CCCG22}' indicates the number of nodes in the converted game of \citet{Carminati21:Public}. 

Column `\textbf{This paper}' reports the time to convergence of the {\em best} CFR variant to an average team exploitability of less than $\eps$ times the range of payoffs of the game. Convergence plots for all CFR variants on all games can be found in the appendix. %
Column `\textbf{ZS22}' reports the time it took ZS22 to compute an equilibrium strategy for team \pmax, to Gurobi's default precision. Finally, column `\textbf{ZFCS22}' reports the time it took ZFCS22 to compute an equilibrium strategy for team \pmax with exploitability of less than $\eps$ times the range of payoffs of the game. The missing values in that column are due to the fact that the implementation of ZFCS22 by the original authors only supported 3-player games.

Overall, our algorithms based on the team belief DAG are generally 2-3 orders of magnitude faster than ZS22. In games with low parameter $k$, our algorithms are also several orders of magnitude faster than ZFCS22, validating the conclusion of \citet{Zhang22:Team}. %
In games with high parameters (\emph{e.g.}, $^3$K8 and $^3$K12), on the other hand, ZFCS22 is significantly more scalable, as it avoids the exponential dependence in the parameters at the cost of requiring the solution to integer programs, for which runtime guarantees are hard to give. Compared to the converted game of \citet{Carminati21:Public}, our team belief DAG is much smaller, often by orders of magnitude, which allows our algorithms to similarly be faster by orders of magnitude. Since \citet{Carminati21:Public} do not give detailed timing results for their implementation for most of the games they tested, we have not included a runtime comparison. However, they reported a runtime of approximately 3 minutes to achieve an exploitability of $0.021$ in $^3$L133, whereas our algorithm took 0.02 seconds to achieve a lower exploitability of $0.001$---a difference of about four orders of magnitude. We believe that some of the difference may be due to their implementation being unoptimized compared to ours, but certainly some of it is not: their converted game is 34x larger than the total size of our DAGs on this game, so we would expect our algorithm to perform approximately that much better with an optimized implementation.
\section{Conclusion and Future Research}
We gave a new representation, the \textit{TB-DAG}, for the decision problem faced by a team of correlating players, which we used to develop new algorithms for solving adversarial team games. Our method enjoys the parameterized complexity bounds of \citet{Zhang22:Team}, and the extensibility and interpretability of \citet{Carminati21:Public}, and ours can be exponentially more efficient in time and space than either and never much less efficient. Experiments showed that modern variants of CFR applied with our TB-DAG give state-of-the-art performance across multiple domains.

This work opens many possible directions for future research, including the following:
\begin{enumerate}
    \item devising a technique to allow the use of {\em Monte Carlo CFR (MCCFR)}~\cite{Lanctot09:Monte} in DAG-form decision problems, and in particular in the TB-DAG;
    \item finding theoretically sound techniques for mitigating the exponential blowup in parameters $w$ and $k$;
    \item finding a ``best-of-both-worlds'' algorithm that combines the strengths of our approach and the single-oracle-based methods;
    \item motivated by the complexity results, investigating whether a practically-fast algorithm exists that uses an integer programming oracle as a subroutine
    \item relaxing the assumption of timeability;
    \item devising a construction that additionally generalizes the {\em triangle-free interaction}~\cite{Farina20:Polynomial}, a known polynomially-solvable subclass of the problem; and
    \item applying other standard game-theoretic techniques in two-player zero-sum games, such as abstraction, dynamic pruning, subgame solving, {\em etc.}, to team games.
\end{enumerate}

\section*{Acknowledgements}
The work of Prof. Sandholm's research group is funded by the National Science Foundation under grants IIS1901403, CCF-1733556, and the ARO under award W911NF2210266.

\bibliographystyle{icml2023}
\bibliography{dairefs}

\newpage
\appendix
\onecolumn

\section{Decision Problems on DAGs}\label{se:dp-dag}

In this section, we discuss how techniques that apply to decision problems, such as CFR can be used on a decision problem that is a DAG, which may be of independent interest beyond team games.
\begin{definition}
    A {\em DAG decision problem} $\mc D =  (\mc H, \mc E)$ is a perfect-information decision problem defined on a DAG, with node set $\mc H$ and edge set $\mc E$, instead of a tree.
\end{definition}
We will insist on the following technical conditions:
\begin{enumerate}
    \item inactive nodes always have exactly one parent;
    \item Nodes along every path alternate between active nodes and inactive nodes; and
    \item If $p_1$ and $p_2$ are two paths from the root ending at the same node, then the last node common to both $p_1$ and $p_2$ is active.
\end{enumerate}
The first two conditions are for expository simplicity and are without loss of generality; the final one is necessary so that the realization form, which we are about to define, makes sense. It is easy to check that our TB-DAG satisfies all three definitions.

The realization form $\vx \in \zo^{\mc Z}$ of a pure strategy is the vector for which $\vx[z] = 1$ if and only if the player plays all the actions on {\em some} $\Root \to z$ path, where condition (3) ensures that there exists {\em at most one} such path. Mixed strategies and their realization forms are then defined analogously to the case of trees.

    {\bf DAG Decision Problems via Scaled Extensions.}
We now show that the set of sequence-form strategies in a DAG can be expressed in terms of {\em scaled extensions}~\cite{Farina19:Efficient}.

\begin{definition}
    Given two nonempty, compact, convex sets $\mc X, \mc Y$ and a linear map $f : \mc X \to \R_{\ge 0}$, the {\em scaled extension} of $\mc X$ with $\mc Y$ via $f$, is defined as
    \begin{align}
        \mc X \se{f} \mc Y = \qty{ (\vx, f(\vx)\vy) : \vx \in \mc X, \vy \in \mc Y}.
    \end{align}
\end{definition}
We now construct the set of sequence-form strategies in a given DAG. We begin with the set $\mc X \gets \{1\}$. Then, for each active node $h$, we perform the operation $\mc X \gets \mc X \se{\vx \mapsto \vx[h]} \Delta^{A_h}$ where
\begin{align}
    \vx[h] := \sum_{(h', h) \in \mc E} \vx[h'].
\end{align}
The restriction of the resulting set $\mc X$ on the set of terminal states $\mc Z$ is exactly the set of sequence-form mixed strategies. Thus, we have shown:
\begin{theorem}\label{th:dag as scaled ext}
    The set of sequence-form strategies on a DAG can be expressed by scaled extension operations with simplices via functions $h : \mc X \to [0, 1]$.
\end{theorem}

\begin{algorithm}[!tb]
    \caption{CFR-based algorithm for DAGs}\label{al:cfr}
    \Fn{\normalfont$\mc D${\sc.NextStrategy}()}{
    \Comment{\color{commentcolor}$\vx'$ will store the {\em unscaled} probabilities]}
    $\vx, \vx' \gets \vec 1_{\in \R^{\mc H}}$\;
    \For{each active node $s$ in $\mc D$ (top down)}{
    \uIf{$s$ is not the root}{
        $\vx[s] \gets \sum_{s' \text{ parent of } s} \vx[s']$\;
    }
    $S \gets \sum_{a \in A_s} R[sa]^+$\;
    \For{each action $a \in A_s$}{
    \uIf{$S = 0$}{
    $\vx'[sa] = 1/\abs{A_s}$
    }
    \Else{
    $\vx'[sa] \gets R[sa]^+/S$
    }
    $\vx[sa] \gets \vx'[sa] \vx[s]$\;
    }
    }
    \Return $\vx$\;
    }
    \Hline{}
    \Fn{\normalfont$\mc D${\sc.ObserveUtility}($\vu \in \R^{\mc Z}$)}{
    \For{each $s \in \mc H \setminus \mc Z$}{
        $\vu[s] \gets 0$
    }
    \For{each active node $s$ in $\mc D$ (bottom up)}{
    $\vu[s] \gets \vu[s] + \sum_{a' \in A_s} \vu[sa'] \vx'[sa']$\;
    \For{each action $s \in A_s$}{
        $R[sa] \gets R[sa] + \vu[sa] - \vu[s]$\;
    }
    \For{each parent $s'$ of $s$}{
        $\vu[s'] \gets \vu[s'] + \vu[s]$\;
    }
    }
    }
\end{algorithm}

{\bf Regret Minimization in DAGs.}\label{sec:dag cfr}
Any set that can be built from scaled extensions and simplexes admits a regret minimizer that can be constructed starting from any simplex regret minimizer~\cite{Farina19:Efficient}. This construction extends CFR~\cite{Zinkevich07:Regret}, and all its modern variants, to such sets. In particular, applying Proposition~1 of \citet{Farina19:Efficient} on top of \Cref{th:dag as scaled ext} gives us:
\begin{corollary}\label{co:cfr complexity}
    CFR can be run on a DAG decision problem $\mc D = (\mc H, \mc E)$, with regret bounded by $O(\abs{\mc H} \sqrt{T})$ after $T$ timesteps and iteration time $O(\abs{{\mc E}})$.
\end{corollary}

Pseudocode for running CFR on an arbitrary DAG can be found in \Cref{al:cfr}. %

\section{The Size of a TB-DAG}\label{se:dag size}

Since all our theoretical results depend on the size of the TB-DAG, it is critical to analyze that size. The hardness result of \citet{Koller92:Complexity} means that our size bounds will not be polynomial. However, we can still bound the size relative to natural parameters related to the complexity of the game.

The correspondence between our construction and that of \citet{Zhang22:Team} allows us to achieve similar theoretical guarantees to that paper. Here, we explicitly give such results in our language. Let $\mc T$ be a team decision problem with node set $\mc S$ and public state set $\mc P$. Define the {\em effective size} of a set of nodes $H \subseteq \mc H$ is the number of distinct team sequences among the nodes in $H$. In all the below theorem statements, $O^*$ hides factors polynomial in the size of the game.

\begin{theorem}\label{th:main}
    The TB-DAG of $\mc T$ has at most $O^*\qty(\sum_{i=1}^w\binom{p}{i}b^w)$ edges%
    , where $p$ is the largest effective size of any public state, $w$ is the largest effective size of any belief, and $b$ is the branching factor of the team decision problem.
\end{theorem}
\begin{proof}
    It suffices to bound the number of inactive nodes, since each inactive node has at most one incoming edge and at most as many outgoing edges as there are public states. A belief $B$ inside a public state $P \in \mc P$, can be uniquely identified by specifying its sequence set $\sigma(B)$ and a node $h$ within it. We have $\abs{\sigma(B)} \le w$ by definition. Hence, there are at most $\sum_{i=1}^w\binom{p}{i}\abs{\mc H}$ beliefs, and at most $b^w$ prescriptions $\va$ at $B$. Multiplying these gives the desired result.
\end{proof}

The parameter $w$ is similar to the namesake parameter in \citet{Zhang22:Team}, except that it might be smaller due to our public observations inducing smaller beliefs. As discussed in that paper, $w$ depends only on the amount of {\em uncommon external information}, that is, {\em observations} (as opposed to decisions by the team) that are not common knowledge to the team.

In a certain family of team decision problems including those with team-public actions, we can do better.
\begin{definition}
    An $n$-player team decision problem is {\em $k$-private} if, in every public state, there are at most $k$ distinct last infosets. That is, $|\{I_i(h) : i \in [n], h \in P\}| \le k$ for every $P \in \mc P$, where $I_i(h)$ is the last infoset reached by player $i$ on the path to $h$, including possibly the infoset containing $h$ itself if $h$ is a decision node of player $i$.
\end{definition}

As an example, in a normal-form game converted to extensive form in the natural manner, $k = n$. This is distinct from the {\em effective size} $p$, which is the total number of {\em team} sequences in $P$. In particular, in games with {\em team-public actions} (such as poker), where each player has at most $t$ private types, we have $k \le nt$.

In a $k$-private team decision problem, it is possible that $w = (k/n)^n$, so \Cref{th:main} gives a bound of $O^*\qty((2bp)^{(k/n)^n})$, which is bad. However, we can improve upon this through a more careful analysis.

\begin{theorem}\label{th:t-private}
    The TB-DAG of a $k$-private team decision problem has at most
    $O^*\qty((b+1)^{k})$ edges.
\end{theorem}

\begin{proof}

    Consider a public state $P$, and let $I$ be any last infoset at $P$. Thus, to specify $I$'s contribution to a belief-prescription pair $B \va$, it suffices to specify one of: either the player does not play to $I$, or the player chooses one of her $b$ possible actions at the sequence. Thus, there are at most $(b+1)^{k}$ possible belief-prescription pairs $B \va$.
\end{proof}

It is possible to ``mix and match'' the analyses of \Cref{th:main,th:t-private} when some public states have low $w$ and some have low $t$. To save the cumbersome notation, we will not do that here.

\subsection{Branching Factor Reduction} \label{se:bf reduction proof}
Since the branching factor $b$ appears as the base of an exponential in \Cref{th:main,th:t-private}, it is natural to ask whether it can reduced without affecting the other parameters. This turns out to be true assuming {\em team-public actions}, which we now formalize.
\begin{definition}
    A team decision problem has {\em team-public actions} if, for all public states $P$ containing active nodes, for all edge labels (\emph{i.e.}, actions or observations) $a \in \bigcup_{h \in P} A_h$, the set $\{ha : h \in P, a \in A_h\}$ is %
    a union of public states.
\end{definition}
Intuitively, this means that any action taken by a team member becomes common knowledge for the team. The definition also allows for information other than the action to become common knowledge, and for some public states to give the team members private information.
\begin{theorem}\label{th:team-public actions}
    Given a team decision problem $\mc T$ with public actions, there exists another realization-equivalent team decision problem $\mc T'$ such that the branching factor of $\mc T'$ is at most $2$ at each active node, the parameters $p, n, k, w$ in $\mc T'$ are the same as in $\mc T$, and the size of the game has increased by at most a polynomial amount.
\end{theorem}

\begin{proof}
    Consider a public state $P$ of $\mc T$. If $P$ contains no active nodes, we leave it alone. Otherwise, let $\mc B$ be an arbitrary binary tree with leaf set $A := \bigcup_{h \in P} A_h$. The internal nodes of $\mc B$ will be labelled with {\em partial actions} $\tilde a$, which we can think of as partial bitstrings of indices of actions in $A$. For each node $h \in P$, we replace $h$ with a modified copy of $\mc B$ wherein subtrees containing no nodes in $A_h$ have been pruned. If $h$ and $h'$ are in the same infoset in $P$, then for every partial action (i.e., nonterminal node) $\tilde a \in \mc B$ we connect $h \tilde a$ and $h' \tilde a$ in an infoset. This creates a new public tree $\mc T'$, whose parameters we must now analyze.

    For each node $h$ of $\mc T$, the construction creates the internal nodes of a public subtree in $\mc T'$ with leaves corresponding to the children of $h$, and $\log \abs{\bigcup_{h \in P} A_h}$ layers, thus introducing at most $\abs{A_h}\log \abs{\bigcup_{h \in P} A_h}$ nodes, so the total number of nodes introduced is $\poly(\abs{\mc H})$ nodes.

    The number of players $n$ remains the same.

    For each new public state $P^*$ constructed in this process, we have $P^* \subseteq P\tilde a := \{h \tilde a : h \in P, \tilde a \preceq a \in A_h\}$ for some $\tilde a$, where $\preceq$ denotes precedence in  $\mc B$ (the subset may not be the whole set, because it is possible for the partial action $\tilde a$ to have already revealed further common knowledge that was not available at $P$). Thus, every team sequence or last infoset in $P$ identifies at most one unique team or last infoset in $P^*$. Namely, a team sequence $\sigma(h)$ in $P$ identifies the team sequence $\sigma(h \tilde a)$ in $P^*$, and $I_i(h)$ identifies the team sequence $I_i(h \tilde a)$. Thus $p$ and $k$ have not increased.

    Finally, for each belief $B \subseteq P$, the largest belief in $P \tilde a$ induced by $B$ is $B \tilde a$, which has no larger effective size. Hence, $w$ has not increased. This completes the proof.
\end{proof}

\begin{corollary}
    In a team decision problem $\mc T$ with public actions, it is possible to create a team DAG for $\mc T$ with $O^*\qty((2p+2)^{w})$ or $O^*(3^k)$ edges.
\end{corollary}

\section{Complexity Results}\label{se:complexity}

For both problems, the goal is to solve the following promise problem: given a two-team zero-sum game $\Gamma$, threshold value $v$, and error $\eps > 0$ (where all the numbers are rational), determine whether the (TMECor or TME) value of the game is $\ge v$, or $< v - \eps$. The allowance of an exponentially-small error is to circumvent issues of bit complexity that arise due to the fact that exact TMEs may not have rational coefficients~\cite{Koller92:Complexity}.

\begin{theorem}[\citealp{Koller92:Complexity,Chu01:NP}]%
    Team-vs-player TMECor and TME are \NP-hard.
\end{theorem}
\begin{proof}
    Given a 3-CNF formula $\phi$ with $m$ clauses and $n$ variables, construct the following game with two players on the maximizing team and no opponent. Nature picks a clause in $\phi$ uniformly at random. Player 1 knows the clause, and selects a variable $x_i$ in the clause. Player 2 learns the variable $x_i$ (but not whether that variable is negated), nor the clause, and selects either true or false.

   It is easy to check that the best possible team value is exactly the maximum fraction of satisfied clauses in any assignment $x$, and this value is achieved when Player 2 plays from that assignment and Player 1 always selects a satisfying literal when one exists. This completes the proof, as by the PCP theorem (e.g.,~\citealp{Haastad01:Some}), approximating the maximum fraction of satisfied clauses in a SAT problem is \NP-hard.
\end{proof}

Since the negation of team-vs-team TMECor is itself and team-vs-player is a special case of team-vs-team, it follows immediately that:
\begin{corollary} %
    Team-vs-team TMECor is \NP-hard and {\sf co-NP}-hard.
\end{corollary}

\begin{theorem}%
    Team-vs-player TMECor is in \NP, and team-vs-team TMECor is in $\Sigma_2^\P \cap \Pi_2^\P$.
\end{theorem}
\begin{proof}
    Given a team-vs-player game $\Gamma$ with $n$ nodes, suppose that its value is $\ge v$. Therefore, there is a distribution $\cal D$ over pure strategies of the team such that no opponent response has value $< v$. By Caratheodory, $\cal D$ is realization equivalent to some distribution $\cal D'$ supported on at most $n$ pure strategies $x_1, \dots, x_n$. Such a distribution can be represented in polynomial bit complexity, and verified in polynomial time (using an IP solver oracle, in the case of team-vs-team).
\end{proof}

\begin{theorem}\label{th:tme-sigma2} %
    Team-vs-player TME is in \NP, and team-vs-team TME is in $\Sigma_2^\P$.
\end{theorem}
\begin{proof}
Consider a TME expressed in behavioral form, that is, for each information set $I$ of the team, we represent a distribution over its actions. Let $\delta > 0$, and consider rounding each entry of the behavioral-form strategy by at most an additive $\delta$ so that the resulting strategy is rational. Let $\vec x'$ be the correlation plan of the resulting strategy. Thus, for any given terminal node $s$, the resulting reach probability $x'[s]$ is perturbed by at most an additive $O(N \delta)$ where $N$ is the number of nodes in the game. Thus, $\norm{\vec x' - \vec x}_1 \le O(N^2 \delta)$. Thus, for any realization-form strategy $\vec y$ for the opponent, we have $\abs{\ip{\vec x' - \vec x, \vec A \vec y}} \le \norm{\vec x' - \vec x}_1 \norm{\vec A \vec y}_\infty \le O(N^2 \delta)$, so $x'$ is $O(N^2 \delta)$-close to the optimal solution. Taking $\delta < O(\eps/N^2)$ thus concludes the proof.
\end{proof}

\begin{theorem}\label{th:tme-sigma2-hard}%
    Team-vs-team TME is $\Sigma_2^\P$-hard.
\end{theorem}
\begin{proof}
    We reduce from $\forall \exists$3-SAT, which is known to be $\Pi_2^\P$-complete~\cite{Schaefer02:Completeness}. The $\forall \exists$3-SAT problem is to, given a $3$-CNF formula $\phi(x, y)$, determine whether $\forall x\ \exists y\ \phi(x, y)$.

    Given a 3-CNF formula $\phi$ with $m$ clauses, $n_1$ variables in $x$, and $n_2$ variables in $y$, construct the following game between the max-team with $2n_1$ players and the min-team with $n_2$ players. Nature chooses a clause $\phi$. For each variable $y_i$ in the clause, Player $i$ on the min-team is asked for an assignment to $y_i$. For each variable $x_i$ in the clause, Players $i$ and $n_1 + i$ on the max-team are asked for an assignment to $x_i$.

    If, for any $x_i$, the two players on the max-team differ in their choice of assignment, the max-team gets value $-M$ where $M$ is a large value. Otherwise, the max-team gets value $1$ if and only if the clause is unsatisfied, else $0$.

    If $\phi$ is not $\forall\exists$-satisfiable, let $x$ be such that $\forall y~\neg \phi(x, y)$, and suppose \pmax plays according to $x$. This forces value at least $1/m$: no matter what pure strategy \pmin plays, there will always exist some clause in $\phi$ that is unsatisfied, so \pmax gets value at least $1/m$. 

The converse will follow, intuitively, from the following observation. For large enough $M$, since \pmax cannot correlate, \pmax's strategy needs to be nearly pure to avoid losing too much utility. Therefore, \pmax must basically fix an assignment $x$. But this cannot achieve large value, because \pmin can simply choose the assignment $y$ that satisfies $\phi$, which makes the value of the game small. We now work through this formally.

\begin{lemma}
Let $x_i$ be a variable. Let $p$ be the probability that Player $i$ plays her less-likely action in a TME. Then $p \le m/M$.
\end{lemma}
\begin{proof}
Variable $x_i$ appears in at least one clause. If that clause is picked by chance (probability $1/m$), then the penalty incurred by the two players is $(M/m)(p(1-q) + q(1-p)) \ge (M/m)(p + q(1/2-p)) \ge (M/m)p$. The result now follows by observing that any strategy incurring penalty greater than $1$ is dominated by a pure strategy.
\end{proof}

For $M$ sufficiently large, then, a TME for \pmax can be rounded to a pure strategy by perturbing each player's probability by at most $m/M$. Suppose $\phi$ is $\forall \exists$-satisfiable. Consider an arbitrary TME for \pmax, and let $x$ be its rounded version---that is, for every variable $i$, \pmax plays from $x$ with probability at least $1 - m/M$. Let $y$ be such that $\phi(x, y)$ is true. The only way for \pmax to get value 1 is for at least one player to play the wrong assignment to at least some variable. By a union bound, this happens with probability at most $mn/M$. Thus, taking $M = 2m^2n$, \pmin ensures that the value of the game is at most $1/(2m)$ by playing from $y$. This completes the proof.
\end{proof}

\begin{theorem}\label{th:tmecor-delta2}
    Team-vs-team TMECor is in $\Delta_2^\P$.
\end{theorem}
\begin{proof}
    Let $\mc X \subset \R^m, \mc Y \subset \R^n$ be the space of realization-form pure strategies of both players, and $\mc A$ be the payoff matrix. Then our goal is to decide whether the polytope
    \begin{align}
        \mc X^* := \qty{ \vx \in \R^m : \quad \mqty{
        \circled{1} & \vx \in \co \mc X,                            \\
        \circled{2} & \vy^\top \mA \vx \le v\ \forall \vy \in \mc Y
            } }
    \end{align}
    is empty. We will show how to separate over $\mc X^*$ with a mixed-integer convex programming oracle, which suffices to complete the proof because such a separating oracle can be used to run the ellipsoid algorithm.

    Given a candidate solution $\vx$, we check both constrants. If $\circled 2$ is violated for some $\vy^* \in \mc Y$, then $\mA \vy^*$ is a separating direction; such $\vy^*$ can be found by an integer programming oracle. If $\circled 1$ is violated, then a separating direction can be found because (strong) separation and optimization are equivalent for well-described polytopes~\cite{Grotschel93:Geometric}, and optimization over $\co \mc X$ is an integer program.
\end{proof}

\begin{theorem}\label{th:tmecor-delta2-hard}
    Team-vs-team TMECor is $\Delta_2^\P$-hard.
\end{theorem}
\begin{proof}
    We reduce from Last-SAT, which is known to be $\Delta_2^\P$-complete~\cite{Krentel88:Complexity}. The Last-SAT problem is to, given a 3-CNF formula $\phi(x)$, decide whether the lexicographically last satisfying assignment of $\phi$ has a $1$ in the least-significant bit.

    Given a 3-CNF formula $\phi$ with $m$ clauses and $n$ variables, we construct the following zero-sum game with $n$ players on each team. First, nature chooses some  $t \in [m+n]$ uniformly.

    If $t \le m$, then let $x_i, x_j, x_k$ be the three variables in clause $t$. Players $i, j, k$ on both teams are asked to assign either true or false to each of the three variables (but are not told anything else). If the max-team satisfies the clause, they score $2m$ points. If the min-team satisfies the clause, they score $1$ point.

    If $t > m$, let $i = t-m$. Both players are asked for their assignments to variable $i$. If Max assigned $1$, then Max scores $2^{-i}$ points. If $i=n$ and Max assigned $1$ then Max scores an additional $2m$ points. If Min assigned $1$, then Min scores $2^{-i}$ points.

    We claim that Max has a mixed strategy scoring $\ge m(2m+1)$, to within error $\eps = 2^{-n}$, if and only if the Last-SAT instance is true. If $\phi$ is not satisfiable, then Max has no way to score $2m^2$ points. So, assume $\phi$ is satisfiable. Let $r(x) \in [0, 1)$ be the value of assignment $x$ when it is expressed as a binary number; i.e., $r(x) = 0.x_1 x_2 \dots, x_n$. Let $x^*$ be the last satisfying assignment.

    If the Last-SAT instance is true, suppose that Max plays according to $x^*$. Then she scores $(m+1)(2m) = m(2m+2)$ points from $t \le m$ and $t = m+n$, and an additional $r(x^*)$ points from $t > m$. But Min has no way to score more than $m+r(x^*)$ points: if she does not play a satisfying assignment then she cannot score more than $m$; if she does, she cannot play one larger than $r(x^*)$. Thus, Max scores at least $m(2m+2)-m = m(2m+1)$ points.

    Conversely, if the Last-SAT instance is false, suppose that Min plays according to $x^*$. Min scores $m + r(x^*)$ points, so Max must score $m(2m+2) + r(x^*)$. But this is impossible: to score $m(2m+2)$, Max must play a satisfying assignment $x$ with $x_n = 1$. But then $r(x) < r(x^*)$ by definition of Last-SAT. This completes the proof.
\end{proof}

\section{Other Omitted Proofs}\label{se:proofs}
\subsection{\Cref{th:dag equivalence}}

We will show the claim for pure strategies, which is enough since mixed and correlated strategies come from taking convex combinations of pure strategies.
\begin{itemize}[left=5mm]
    \item[$(\Rightarrow)$] Consider a pure correlation plan $\vx$. Consider the pure strategy in the TB-DAG in which the team chooses the prescription in each belief consistent with $\vx$, inducing a TB-DAG-form strategy $\vx'$.

        Let $z$ be a terminal node in $\mc T$, and suppose $\vx[z] = 1$. We need to demonstrate a path through $\mc D$ leading to $\{z\}$ such that $\vx$ plays every action prescribed along that path. Consider the path through $\mc D$ defined by following the prescriptions of $\vx$, and always selecting the public observation that leads to $z$. By construction of the TB-DAG, this path  must end exactly at $z$, so $\vx'[\{z\}] = 1$.

        Conversely, suppose that such a path exists. Then, every infoset $I \preceq z$ must have appeared in exactly one belief node $B$ along the path, and, at that belief node, in order for $\{z\}$ to still have been reachable, the team must have chosen the action at $I$ leading to $z$. Thus, the team plays all actions on the path from the root to $z$, so $\vx[z] = 1$.
    \item[$(\Leftarrow)$] Consider a pure strategy $\vx'$ in $\mc D$, and let $\vx'$ be its realization form. Define the pure strategy $\vx$ in $\mc T$ as follows. In each level $\mc H^\ell$ of $\mc T$, the strategy $\vx'$ induces a collection of disjoint beliefs $\mc B^\ell(\vx')$. For each such belief $B \in \mc B^\ell(\vx')$, let $\va$ be the prescription in $\vx'$ at $B_k$. At every infoset $I$ intersecting $B_k$, define $\vx$ to play $a_I$ at $I$. This strategy is well-defined because no two beliefs $B, B' \in \mc B^\ell(\vx')$ can intersect the same infoset (otherwise they would not be distinct beliefs!), and if we have not defined an action at an infoset, that means $\vx'$ plays to no node in that infoset.  We claim that $\vx$ defined in this way is realization-equivalent to $\vx'$.

        Suppose $\vx'[\{z\}] = 1$; that is, there is a path through $\mc D$ ending at $\{z\}$ at which $\vx'$ plays every prescription. Then, at each belief $B$ along this path, if $B$ contains an infoset $I \preceq z$, then $I \cap B \ne \Root$. Thus, the team plays all actions on the root $\to z$ path, so $\vx[z] = 1$.

        Conversely, suppose $\vx[z] = 1$. Then we construct a path through a DAG that follows the prescriptions of $\vx'$ and always selects the public observation leading to $z$. By induction, such a path must always contain an ancestor of $z$; in particular, once it reaches the layer of $z$, it must have reached $\{z\}$.
\end{itemize}

\subsection{\Cref{pr:public state comparison}}

Let $B$ be any belief in the public state TB-DAG. In the (non-public-state) TB-DAG, $B$ splits into disjoint beliefs $B_1, \dots, B_m$. Let $A_1, \dots, A_m$ be the sizes of the prescription spaces at $B_1, \dots, B_m$ respectively. Then $B$ has $A_1A_2 \dots A_m$ children, so $B$ induces $1 + A_1A_2 \dots A_m$ nodes in the public state TB-DAG. On the other hand, the beliefs $B_1, \dots, B_m$ in the TB-DAG will have $A_1, \dots, A_m$ children respectively, accounting for a total of $m + A_1 + \dots + A_m \le 2m A_1 \dots A_m$ nodes. Now observing simply that $m \le p$ completes the proof.

\subsection{\Cref{pr:inflate1}}

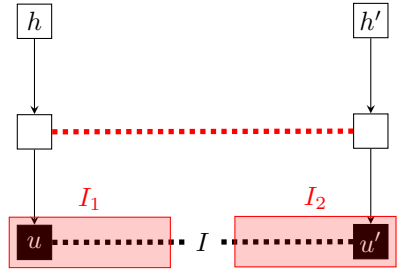
\begin{figure}[t]
\tikzset{
    every path/.style={-stealth},
    every node/.style={draw=black, fill=white, inner sep=0pt, minimum size=1.3em},
    pubnode/.style={fill=red, fill opacity=0.2, draw=red, inner sep=0.1cm},
    infoset/.style={-, dotted, line width=2pt},
    circled/.style={draw, circle, minimum size=1.3em, inner sep=0mm},
}
\newcommand{\mymidrulegray}{\arrayrulecolor{gray}\midrule}
\forestset{
    dec/.style={draw, fill=black,text=white},
    terminal/.style={}, %
    p1/.style={regular
        polygon, regular polygon
        sides=3, inner sep=2pt, fill=p1color, draw=none},
    p2/.style={p1, shape border rotate=180, fill=p2color},
    parent/.style={no edge,tikz={\draw (#1.south) -- (!.north); }},
}
    \centering
\begin{forest}
for tree={parent anchor=south, child anchor=north, s sep=4cm, l sep=1cm}
[,draw=none
    [$h$,no edge
        [,name=x1 [ $u$,name=u1,dec
        ]]
    ]
    [$h'$,no edge
        [,name=x2 [ $u'$,name=u2,dec
        ]]
    ]
]
\draw[infoset] (u1) -- (u2) node[midway,draw=none,fill=white]{$I$};
\draw[infoset,draw=red] (x1) -- (x2);
\node[draw=none, fill=none, right=1cm of u1] (u1x) {};
\node[pubnode, fit=(u1)(u1x), label=above:{\textcolor{red}{$I_1$}}] {};
\node[draw=none, fill=none, left=1cm of u2] (u2x) {};
\node[pubnode, fit=(u2)(u2x), label=above:{\textcolor{red}{$I_2$}}] {};
\end{forest}
\caption{A pictoral representation of the proof of \Cref{pr:inflate1}. Since $h$ and $h'$ can be played simultaneously but $u$ and $u'$ cannot, there must be an infoset like the red dotted one connecting a child of $h$ to a child of $h'$. Therefore, inflation cannot break existing edges between played nodes.}
\end{figure}

Let $I = I_1 \sqcup I_2$ be an inflatable infoset. The only place where inflation can have an effect is the construction of the public observations $P_i$. Hence, let $O$ be inactive, and $h, h' \in O$. We need to show that inflating cannot remove an $(h, h')$ edge in $G[O]$. Suppose it did. Then (WLOG) let us say that $h \preceq u \in I_1$ and $h' \preceq u' \in I_2$. But $O$ is a valid node in $\mc D$, so it is possible for the team to play to both nodes $h$ and $h'$ simultaneously. But then there must be an infoset connecting some node on the $h \to u$ path to some node on the $h' \to u'$ path---otherwise, it would be possible for the team to play to both $u$ and $u'$ simultaneously, which violates inflatability of $I$. This completes the proof. \qedhere

\subsection{\Cref{pr:inflate2}}

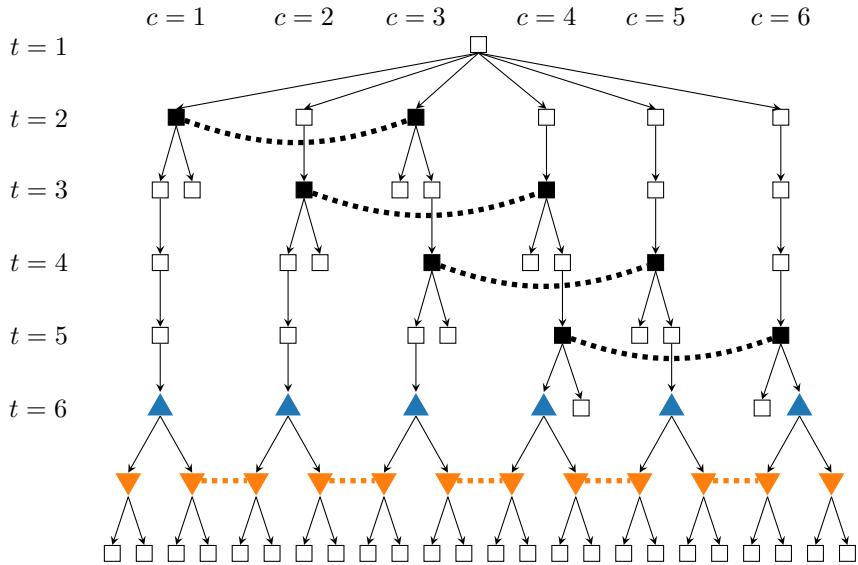
\begin{figure}[t]
\tikzset{
    every path/.style={-stealth},
    every node/.style={draw=black, fill=white, inner sep=3pt},
    pubnode/.style={fill=red, fill opacity=0.2, draw=red},
    infoset/.style={-, dotted, line width=2pt},
    circled/.style={draw, circle, minimum size=1.3em, inner sep=0mm},
}
\forestset{
    dec/.style={draw, fill=black, inner sep=3pt},
    terminal/.style={}, %
    p1/.style={regular
        polygon, regular polygon
        sides=3, inner sep=2pt, fill=p1color, draw=none},
    p2/.style={p1, shape border rotate=180, fill=p2color},
}
    \centering
\begin{forest}
for tree={parent anchor=south, child anchor=north, s sep=0.2cm, l sep=0.5cm}
[
    [,dec,name=A1
        [,name=y3 [,name=y4 [,name=y5 [,p1,name=y6
            [,p2 [] []] 
            [,p2,name=1 [] []]
        ]]]]
        [,terminal]
    ]
    [,name=x2 [,dec,name=A2
        [[[,p1
                [,p2,name=2 [] []] 
                [,p2,name=3 [] []]
        ]]]
        [,terminal]
    ]]
    [,dec,name=B1
        [,terminal]
        [[,dec,name=A3
            [[,p1
                [,p2,name=4 [] []] 
                [,p2,name=5 [] []]
            ]]
            [,terminal]
        ]]
    ]
    [,name=x4 [,dec,name=B2
        [,terminal]
        [[,dec,name=A4
            [,p1
                [,p2,name=6 [] []] 
                [,p2,name=7 [] []]
            ]
            [,terminal]
        ]]
    ]]
    [,name=x5 [[,dec,name=B3
        [,terminal]
        [[,p1
                [,p2,name=8 [] []] 
                [,p2,name=9 [] []]
        ]]
    ]]]
    [,name=x6 [[[,dec,name=B4
        [,terminal]
        [,p1
                [,p2,name=10 [] []] 
                [,p2,name=11 [] []]
        ]
    ]]]]
]
\draw[infoset, bend right=20] (A1) to (B1);
\draw[infoset, bend right=20] (A2) to (B2);
\draw[infoset, bend right=20] (A3) to (B3);
\draw[infoset, bend right=20] (A4) to (B4);
\draw[infoset, draw=p2color] (1) to (2);
\draw[infoset, draw=p2color] (3) to (4);
\draw[infoset, draw=p2color] (5) to (6);
\draw[infoset, draw=p2color] (7) to (8);
\draw[infoset, draw=p2color] (9) to (10);
\node[above=1cm of A1,draw=none]{$c=1$};
\node[above=1cm of x2,draw=none]{$c=2$};
\node[above=1cm of B1,draw=none]{$c=3$};
\node[above=1cm of x4,draw=none]{$c=4$};
\node[above=1cm of x5,draw=none]{$c=5$};
\node[above=1cm of x6,draw=none]{$c=6$};
\node[left=1cm of y3,draw=none](t3label){$t=3$};
\node[above=0.5cm of t3label,draw=none](t2label){$t=2$};
\node[above=0.5cm of t2label,draw=none](t1label){$t=1$};
\node[left=1cm of y4,draw=none]{$t=4$};
\node[left=1cm of y5,draw=none]{$t=5$};
\node[left=1cm of y6,draw=none]{$t=6$};
\end{forest}
\caption{The counterexample for \Cref{pr:inflate2}, for $C=6$. All solid-colored nodes (\pone, \ptwo, and $\blacksquare$) are active (we split them into three different symbols and types so that we can discuss each one separately).}%
    \label{fi:counterexample}
\end{figure}

The counterexample in \Cref{fi:inflation-counterexample} would work if it were not for the fact that all of \ptwo's infosets inflate. Therefore, for our proof of this result, we use a similar gadget at the bottom of the game, but  ensure that \ptwo's infosets do not inflate.

Consider the following family of team decision problems, parameterized by a integer $C > 1$. We will not distinguish the players on the team except for \pone and \ptwo. First, nature picks an integer $c \in \{ 1, \dots, C\}$. Over the next $C-2$ layers $t=2, 3, \dots, C-1$, if $c \in \{ t-1, t+1 \}$, a player who cannot distinguish the two cases chooses an action $a \in \{ -1, +1 \}$. If $c = t+a$, then the game continues; otherwise, the game ends.

Finally, player \pone, who has perfect information about $c$ chooses either $c$ or $c+1$. Then, player \ptwo, observing \pone's action but {\em not} the value $c$, picks one of two options.

The resulting team decision problem is visualized in \Cref{fi:counterexample}. We observe the following things about it.

\begin{enumerate}
    \item No infoset inflates: all nontrivial infosets have size $2$, and it is easy to check that for all such infosets it is always possible to play to both nodes in them. This is in stark contrast to the earlier counterexample, in which inflation was enough to achieve a small representation.
    \item Every \pone-node in layer $C$ is in the same public state, and it is always possible to play to at least $C/2$ of them. Therefore, even if one runs inflation beforehand (which does nothing), if using public-state-based beliefs, there will be a belief with $2^{C/2}$ prescriptions. Thus, the public-state-based team belief DAG, and also the construction of \citet{Zhang22:Team}, will have size at least $2^{C/2}$.
\end{enumerate}

We now claim that any given (nonterminal) node takes part in $O(1)$ public observation-based beliefs, and such beliefs never touch more than $O(1)$ different infosets. This would complete the proof, because this would mean that the team belief DAG has size at most $O(C^2)$. Clearly, the claim is true at the final layer, because each infoset contains only at most two nodes.

Fix a layer $t \in \{ 2, \dots, C \}$, and number the nonterminal nodes in it according to the nature choice $c$.
The induced connectivity subgraph $G[{\mc H}_t]$, where ${\mc H}_t$ is the set of nodes at level $t$, has edges linking $j$ to $j+1$ for all $j$, as well as edges between $j$ and $j+2$ whenever $j \ge t-1$. Further, every node $j > t$ must be played to, because there are no player nodes on the path from root to such nodes.

Let $B$ be a belief containing node $i \le t$. (If no such $i$ exists, then since we play to every node $j > t$, the belief $B$ must be exactly $\{ t+1, \dots, C\}$ which touches exactly one infoset) We claim that specifying which one of $i-1$ or $i+1$ is in $B$ is enough to fully determine $B$.

By construction, since we play to $i$, it is impossible to also play to node $i-2$, because they were linked by an infoset at time $i-1 < t$. Thus, any node $j < i-2$ cannot be part of the same belief, because there is no infoset linking any such node to any node $j' \ge i-1$ except the one connecting a descendant of $i-1$ to a descendant of $i-2$. We now consider cases.
\begin{enumerate}
    \item If $i \le t-2$, then the same argument applies to node $i+2$, so there can be at most $O(1)$ nodes in the belief.
    \item If $i \ge t-1$, then $i$ is connected to every node $j>t$, and all are played. Thus, it is only a question of whether the node $t$ itself is played, but in any case, once again there can be at most $O(1)$ nodes in the belief.
\end{enumerate}
Thus, the total number of beliefs is at most $O(C^2)$, and each has $O(1)$ branching factor, so the team belief DAG is also of size $O(C^2)$. This is exponentially smaller than the public-belief-based team belief DAG or the construction of \citet{Zhang22:Team}. \qed

A practical experiment backs up these results. When $C = 16$, using public observations generates a DAG with around 1000 edges; using public states generates a DAG with 30 million edges.

    {\em Remark.}
\Cref{pr:inflate1,pr:inflate2} are mostly of theoretical interest. Their impact on practical game instances is negligible. Instead, the practical improvements in \Cref{se:optimizations} have significantly more effect in practice.

\section{Example in which the TB-DAG is exponentially better than the expanded game tree}\label{se:carminati counterexample}

In this section, we exhibit another explicit counterexample in which our TB-DAG will be exponentially smaller than the converted game of \citet{Carminati21:Public}. Since that converted game effectively has (at least) one node corresponding to every path from the root in our TB-DAG, it suffices to exhibit a game in which our TB-DAG has exponentially many paths. Consider the tree from \Cref{fi:example-game}, duplicated several times by repeatedly attaching a copy of itself at node {\sf H}. Let the tree with $n$ such duplicates be denoted $\mc T_n$ (so that $\mc T_1$ is the original game). In $\mc T_1$, there are two paths to the active node {\sf H}. Following the pattern, in $\mc T_n$, there will be $2^n$ paths to the last copy of node {\sf H}. Thus, the TB-DAG of $\mc T_n$ will have $O(n)$ nodes but $2^{O(n)}$ paths, which is what we wanted to show.

\section{TB-DAG Postprocessing Techniques }\label{se:optimizations}
In practice, the construction of \Cref{al:team-dag} is suboptimal in several ways. These do not affect the theoretical statements as the primary focus of those is isolating the dependency on our parameters of interest, but they can significantly affect the practical performance, so we apply them in the experiments.
\begin{enumerate}
    \item If two terminal nodes $z, z'$ correspond to the same team sequence, we remove one of them (say, $z'$) from our DAG because it is redundant, and alias $\vx[\{z'\}]$ to $\vx[\{z\}]$. If this removal causes a section of the DAG to no longer contain any terminal children, we remove that section as well.
    \item If an active node has at most one parent and at most one child, we remove it and its child, and connect its parent directly to its grandchildren.
\end{enumerate}
In particular, if the team has perfect recall, the above two optimizations are sufficient for the team belief DAG to coincide with the sequence form.

\Cref{app:additional experiments} empirically investigates the benefit of postprocessing the TB-DAG.

\section{Additional Experimental Details}
\label{app:additional experiments}

\subsection{TB-DAG Construction Time}
\Cref{tab:construction times} shows the construction times for the TB-DAG, the linear programming approach of \citet{Zhang22:Team} (ZS22), and von Stengel-Forges polytope for the column-generation technique of \citet{Zhang22:Optimal} on the left part, as well as a version of our main results table with these construction times added and also including LP running on the TB-DAG on the right part. It is worth mentioning that our implementation of the von Stengel-Forges polytope construction is highly optimized, whereas the TB-DAG and ZS22 constructions are not. In particular, the latter two are not parallelized, and parallelization would easily save a factor of approximately the number of threads.

The results show that our algorithm, even when considering the unoptimized TB-DAG construction time, is significantly faster than ZS22 and column generation. In other words, the conclusions reported in the paper do not significantly change.

\begin{table}[H]
\centering
\colorlet{teamcol}{gray}
\newcommand{\cbox}[2]{\fcolorbox{white}[rgb]{1,1,1}{\phantom{28m 13s}\llap{#2}}}
\centering
\newcommand{\mymidrulegray}{\arrayrulecolor{gray}\mymidrule}
\newcommand{\mymidrule}{
\cmidrule(lr){1-5}
\cmidrule(lr){6-7}
\cmidrule(lr){8-9}
\cmidrule(lr){10-10}
\cmidrule(lr){11-15}
\arrayrulecolor{black}}
\newcommand{\unk}{\textcolor{black!30}{---}}
\setlength{\tabcolsep}{2mm}
\newcommand{\teamprint}[1]{\{\textcolor{teamcol}{#1}\}}
\scalebox{.9}{\begin{tabular}{lrrr|rrrr}
\toprule
      \multirow{2}{*}{\makebox[15mm][l]{{\bf Game} $\{\pmin\}$}} &
      \bf TB-DAG &
      \bf ZS22 LP &
      \bf CG VSF &
      \bf CFR (Ours) &
      \bf LP (Ours) &
      \bf LP (ZS22) &
      \bf CG
      \\
      & constr. time
      & constr. time
      & constr. time
      & $\epsilon = 10^{-3}$
      & 
      &
      & $\epsilon = 10^{-3}$
      \\
\midrule
    $^3$K3 \teamprint{3} &  0.00s &  0.00s & 0.00s &                \cbox{0.7843137254901961,0.7843137254901961,1.0}{0.00s} &                 \cbox{0.7843137254901961,0.7843137254901961,1.0}{0.00s} &  \cbox{0.8126105344098423,0.8126105344098423,0.9893886966551326}{0.01s} &                \cbox{0.7843137254901961,0.7843137254901961,1.0}{0.00s} \\
    $^3$K4 \teamprint{3} &  0.01s &  0.01s & 0.00s & \cbox{0.8199923106497501,0.8199923106497501,0.9866205305651672}{0.01s} &  \cbox{0.8913494809688581,0.8913494809688581,0.9598615916955017}{0.02s} &  \cbox{0.9460207612456747,0.9282583621683967,0.9282583621683967}{0.03s} &                \cbox{0.7843137254901961,0.7843137254901961,1.0}{0.01s} \\
    $^3$K6 \teamprint{3} &  1.03s &  1.03s & 0.00s &  \cbox{0.984313725490196,0.8261437908496732,0.8261437908496732}{1.05s} &                 \cbox{1.0,0.7843137254901961,0.7843137254901961}{1.53s} &                 \cbox{1.0,0.7843137254901961,0.7843137254901961}{2.24s} &                \cbox{0.7843137254901961,0.7843137254901961,1.0}{0.14s} \\
    $^3$K8 \teamprint{3} &  1m 6s & 1m 25s & 0.01s &               \cbox{1.0,0.7843137254901961,0.7843137254901961}{1m 11s} &                \cbox{1.0,0.7843137254901961,0.7843137254901961}{1m 47s} &                \cbox{1.0,0.7843137254901961,0.7843137254901961}{4m 23s} &                \cbox{0.7843137254901961,0.7843137254901961,1.0}{0.24s} \\
  $^4$K5 \teamprint{3,4} &  0.55s &  0.30s &  \unk &                \cbox{0.7843137254901961,0.7843137254901961,1.0}{0.58s} &  \cbox{0.8864282968089197,0.8864282968089197,0.9617070357554787}{1.22s} &  \cbox{0.8704344482891195,0.8704344482891195,0.9677047289504037}{1.09s} &                                                                   \unk \\
    $^4$K5 \teamprint{4} & 13.71s & 27.62s &  \unk &               \cbox{0.7843137254901961,0.7843137254901961,1.0}{15.30s} & \cbox{0.9769319492502884,0.8458285274894272,0.8458285274894272}{1m 36s} &                \cbox{1.0,0.7843137254901961,0.7843137254901961}{3m 49s} &                                                                   \unk \\
  $^3$L133 \teamprint{3} &  0.49s &  0.13s & 0.02s &                \cbox{0.7843137254901961,0.7843137254901961,1.0}{0.51s} &  \cbox{0.8335255670895809,0.8335255670895809,0.9815455594002307}{0.73s} &  \cbox{0.8113802383698577,0.8113802383698577,0.9898500576701269}{0.63s} &               \cbox{1.0,0.7843137254901961,0.7843137254901961}{24.91s} \\
  $^3$L143 \teamprint{3} &  1.39s &  0.99s & 0.05s &                \cbox{0.7843137254901961,0.7843137254901961,1.0}{1.49s} &  \cbox{0.9510957324106113,0.9147251057285659,0.9147251057285659}{5.73s} &    \cbox{0.9718569780853518,0.859361783929258,0.859361783929258}{8.54s} &                \cbox{1.0,0.7843137254901961,0.7843137254901961}{2m 5s} \\
  $^3$L151 \teamprint{3} &  1.54s &  1.14s & 0.04s &                \cbox{0.7843137254901961,0.7843137254901961,1.0}{1.73s} &  \cbox{0.9510957324106113,0.9147251057285659,0.9147251057285659}{6.63s} & \cbox{0.9746251441753172,0.8519800076893502,0.8519800076893502}{10.46s} &  \cbox{0.8630526720492118,0.8630526720492118,0.970472895040369}{3.10s} \\
  $^3$L153 \teamprint{3} & 16.03s & 11.52s & 0.12s &               \cbox{0.7843137254901961,0.7843137254901961,1.0}{17.27s} &   \cbox{0.9963091118800461,0.794156093810073,0.794156093810073}{2m 40s} &                \cbox{1.0,0.7843137254901961,0.7843137254901961}{4m 30s} &               \cbox{1.0,0.7843137254901961,0.7843137254901961}{7m 23s} \\
  $^3$L223 \teamprint{3} &  0.13s &  0.19s & 0.05s &                \cbox{0.7843137254901961,0.7843137254901961,1.0}{0.17s} &  \cbox{0.8482891195693963,0.8482891195693963,0.9760092272202999}{0.26s} &  \cbox{0.9257977700884275,0.9257977700884275,0.9469434832756632}{0.46s} &               \cbox{1.0,0.7843137254901961,0.7843137254901961}{13.54s} \\
  $^3$L523 \teamprint{3} & 18.02s & 30.47s & 6.83s &               \cbox{0.7843137254901961,0.7843137254901961,1.0}{29.28s} &   \cbox{0.859361783929258,0.859361783929258,0.9718569780853518}{51.00s} &  \cbox{0.970472895040369,0.8630526720492118,0.8630526720492118}{2m 43s} &               \cbox{1.0,0.7843137254901961,0.7843137254901961}{5h 35m} \\
$^4$L133 \teamprint{3,4} &  2.03s &  1.10s &  \unk &                \cbox{0.7843137254901961,0.7843137254901961,1.0}{2.24s} &  \cbox{0.9441753171856978,0.9331795463283352,0.9331795463283352}{7.53s} &  \cbox{0.9492502883506344,0.9196462898885044,0.9196462898885044}{8.27s} &                                                                   \unk \\
    $^3$D3 \teamprint{3} &  0.80s &  0.64s & 0.09s &                \cbox{0.7843137254901961,0.7843137254901961,1.0}{0.91s} &   \cbox{0.8581314878892734,0.8581314878892734,0.972318339100346}{1.57s} &  \cbox{0.9344098423683198,0.9344098423683198,0.9437139561707035}{2.75s} &               \cbox{1.0,0.7843137254901961,0.7843137254901961}{11.13s} \\
    $^3$D4 \teamprint{3} &  1m 3s & 43.39s & 1.57s &               \cbox{0.7843137254901961,0.7843137254901961,1.0}{1m 25s} & \cbox{0.9423298731257209,0.9381007304882737,0.9381007304882737}{4m 36s} &  \cbox{0.9764705882352941,0.8470588235294118,0.8470588235294118}{9m 1s} &               \cbox{1.0,0.7843137254901961,0.7843137254901961}{3h 19m} \\
  $^4$D3 \teamprint{2,4} & 27.05s & 10.86s &  \unk &               \cbox{0.7843137254901961,0.7843137254901961,1.0}{29.36s} & \cbox{0.8433679354094579,0.8433679354094579,0.9778546712802768}{45.52s} & \cbox{0.9455594002306805,0.9294886582083813,0.9294886582083813}{1m 41s} &                                                                   \unk \\
$^6$D2 \teamprint{2,4,6} & 10.74s &  6.46s &  \unk &               \cbox{0.7843137254901961,0.7843137254901961,1.0}{12.45s} & \cbox{0.8027681660899654,0.8027681660899654,0.9930795847750865}{14.33s} & \cbox{0.8679738562091504,0.8679738562091504,0.9686274509803922}{22.99s} &                                                                   \unk \\
  $^6$D2 \teamprint{4,6} & 16.55s & 12.10s &  \unk &               \cbox{0.7843137254901961,0.7843137254901961,1.0}{20.36s} & \cbox{0.8409073433294887,0.8409073433294887,0.9787773933102653}{30.92s} & \cbox{0.9672433679354094,0.8716647443291041,0.8716647443291041}{1m 46s} &                                                                   \unk \\
    $^6$D2 \teamprint{6} & 31.00s & 37.05s &  \unk & \cbox{0.7990772779700115,0.7990772779700115,0.9944636678200692}{1m 1s} &                \cbox{0.7843137254901961,0.7843137254901961,1.0}{54.89s} &  \cbox{0.9995386389850057,0.7855440215301807,0.7855440215301807}{9m 0s} &                                                                   \unk \\
\bottomrule
\end{tabular}}
\caption{(Left) Comparison of construction times for our TB-DAG, the LP of \citet{Zhang22:Team}, and the von Stengel-Forges polytope-based column-generation (CG) technique of \cite{Zhang22:Optimal}. (Right) Cumulative running times including construction times for the different algorithms benchmarked in the paper.}
\label{tab:construction times}
\end{table}

\subsection{Effect of Postprocessing Step on Final TB-DAG Size}

\Cref{tab:postproc size effect} shows the TB-DAG size (number of edges) with and without the practical tricks. The results show that postprocessing significantly reduces the size of the TB-DAG, sometimes by a factor of 10 or more. Therefore, this ablation confirms that postprocessing of the DAG is an important step in the algorithm.

\begin{table}[H]
\sisetup{group-separator = {,}, group-minimum-digits=3}
\colorlet{teamcol}{gray}
\newcommand{\cbox}[2]{\fcolorbox{white}[rgb]{1,1,1}{\phantom{28m 13s}\llap{#2}}}
\centering
\newcommand{\mymidrulegray}{\arrayrulecolor{gray}\mymidrule}
\newcommand{\mymidrule}{
\cmidrule(lr){1-5}
\cmidrule(lr){6-7}
\cmidrule(lr){8-9}
\cmidrule(lr){10-10}
\cmidrule(lr){11-15}
\arrayrulecolor{black}}
\newcommand{\unk}{\textcolor{black!30}{---}}
\setlength{\tabcolsep}{2mm}
\newcommand{\teamprint}[1]{\{\textcolor{teamcol}{#1}\}}
\scalebox{.95}{\begin{tabular}{l|rrr|rrr}
\toprule
      \multirow{2}{*}{\makebox[15mm][l]{{\bf Game} $\{\pmin\}$}} &
      \multicolumn{3}{c|}{\bf Team \pmax's DAG size (num. edges)} &
      \multicolumn{3}{c}{\bf Team \pmin's DAG size (num. edges)}
      \\
      & Postprocessing
      & No postprocessing
      & Ratio
      & Postprocessing
      & No postprocessing
      & Ratio
      \\
\midrule
   $^3$K3 \teamprint{3} &      \num{918} &      \num{976} &  1.06 &      \num{36} &     \num{164} &   4.56 \\
    $^3$K4 \teamprint{3} &     \num{6711} &     \num{7143} &  1.06 &      \num{48} &     \num{378} &   7.88 \\
    $^3$K6 \teamprint{3} &   \num{336944} &   \num{346906} &  1.03 &      \num{72} &    \num{1406} &  19.53 \\
    $^3$K8 \teamprint{3} & \num{15564765} & \num{15725273} &  1.01 &      \num{96} &    \num{3634} &  37.85 \\
  $^4$K5 \teamprint{3,4} &   \num{124875} &   \num{136403} &  1.09 &   \num{15415} &   \num{22323} &   1.45 \\
    $^4$K5 \teamprint{4} &  \num{4658070} &  \num{4743796} &  1.02 &     \num{120} &    \num{4432} &  36.93 \\
  $^3$L133 \teamprint{3} &    \num{49005} &    \num{54101} &  1.10 &     \num{684} &    \num{7106} &  10.39 \\
  $^3$L143 \teamprint{3} &   \num{417027} &   \num{444299} &  1.07 &    \num{1200} &   \num{19914} &  16.59 \\
  $^3$L151 \teamprint{3} &   \num{496196} &   \num{512546} &  1.03 &    \num{1500} &   \num{12292} &   8.19 \\
  $^3$L153 \teamprint{3} &  \num{3486091} &  \num{3609851} &  1.04 &    \num{1860} &   \num{45682} &  24.56 \\
  $^3$L223 \teamprint{3} &    \num{45913} &    \num{76157} &  1.66 &    \num{2436} &   \num{31734} &  13.03 \\
  $^3$L523 \teamprint{3} &  \num{4183685} &  \num{6505495} &  1.55 &  \num{220704} & \num{2521646} &  11.43 \\
$^4$L133 \teamprint{3,4} &   \num{158058} &   \num{251742} &  1.59 &  \num{155475} &  \num{251895} &   1.62 \\
    $^3$D3 \teamprint{3} &   \num{215967} &   \num{435475} &  2.02 &    \num{1521} &   \num{76189} &  50.09 \\
    $^3$D4 \teamprint{3} & \num{13749608} & \num{32130090} &  2.34 &   \num{16380} & \num{1646112} & 100.50 \\
  $^4$D3 \teamprint{2,4} &  \num{1217310} &  \num{6731374} &  5.53 & \num{1155144} & \num{6523226} &   5.65 \\
$^6$D2 \teamprint{2,4,6} &   \num{457795} &  \num{3194399} &  6.98 &  \num{389995} & \num{2981963} &   7.65 \\
  $^6$D2 \teamprint{4,6} &  \num{2029546} &  \num{6385216} &  3.15 &   \num{60717} & \num{2014433} &  33.18 \\
    $^6$D2 \teamprint{6} &  \num{7395885} & \num{13947463} &  1.89 &    \num{5550} & \num{1941206} & 349.77 \\
\bottomrule
\end{tabular}}
    \caption{Comparison of TB-DAG size (number of edges) with and without TB-DAG postprocessing (\Cref{se:optimizations}).}
    \label{tab:postproc size effect}
\end{table}

\subsection{Effect of Postprocessing Step on Solver Performance}

\Cref{tab:postproc runtime effect} compares the performance of our DAG-form generalization of CFR applied on the DAG produced with and without postprocessing. Again, we see a benefit associated with postprocessing.

\begin{table}[H]
\sisetup{group-separator = {,}, group-minimum-digits=3}
\colorlet{teamcol}{gray}
\newcommand{\cbox}[2]{\fcolorbox{white}[rgb]{1,1,1}{\phantom{28m 13s}\llap{#2}}}
\centering
\newcommand{\mymidrulegray}{\arrayrulecolor{gray}\mymidrule}
\newcommand{\mymidrule}{
\cmidrule(lr){1-5}
\cmidrule(lr){6-7}
\cmidrule(lr){8-9}
\cmidrule(lr){10-10}
\cmidrule(lr){11-15}
\arrayrulecolor{black}}
\newcommand{\unk}{\textcolor{black!30}{---}}
\setlength{\tabcolsep}{2mm}
\newcommand{\teamprint}[1]{\{\textcolor{teamcol}{#1}\}}
\scalebox{.95}{\begin{tabular}{l|rrr|rrr}
\toprule
      \multirow{2}{*}{\makebox[15mm][l]{{\bf Game} $\{\pmin\}$}} &
      \multicolumn{3}{c|}{\bf Approximation $\epsilon = 10^{-3}$} &
      \multicolumn{3}{c}{\bf Approximation $\epsilon = 10^{-4}$}
      \\
      & CFR (postproc)
      & CFR (no postproc.)
      & Ratio
      & CFR (postproc)
      & CFR (no postproc.)
      & Ratio
      \\
\midrule
    $^3$K3 \teamprint{3} &  0.00s &  0.00s & 1.00 &  0.00s &  0.00s & 1.00 \\
    $^3$K4 \teamprint{3} &  0.00s &  0.00s & 1.00 &  0.00s &  0.00s & 1.00 \\
    $^3$K6 \teamprint{3} &  0.03s &  0.03s & 1.19 &  0.12s &  0.15s & 1.20 \\
    $^3$K8 \teamprint{3} &  4.73s &  4.44s & 0.94 & 32.36s & 27.45s & 0.85 \\
  $^4$K5 \teamprint{3,4} &  0.03s &  0.03s & 1.23 &  0.05s &  0.06s & 1.21 \\
    $^4$K5 \teamprint{4} &  1.59s &  1.96s & 1.23 &  6.34s &  7.98s & 1.26 \\
  $^3$L133 \teamprint{3} &  0.02s &  0.02s & 1.18 &  0.05s &  0.06s & 1.22 \\
  $^3$L143 \teamprint{3} &  0.10s &  0.17s & 1.67 &  0.48s &  0.75s & 1.57 \\
  $^3$L151 \teamprint{3} &  0.18s &  0.21s & 1.16 &  0.50s &  0.63s & 1.25 \\
  $^3$L153 \teamprint{3} &  1.24s &  1.42s & 1.14 &  4.94s &  9.34s & 1.89 \\
  $^3$L223 \teamprint{3} &  0.03s &  0.09s & 2.74 &  0.08s &  0.21s & 2.73 \\
  $^3$L523 \teamprint{3} & 11.26s & 28.18s & 2.50 & 24.86s & 58.75s & 2.36 \\
$^4$L133 \teamprint{3,4} &  0.21s &  0.42s & 2.01 &  0.92s &  1.79s & 1.95 \\
    $^3$D3 \teamprint{3} &  0.11s &  0.66s & 5.85 &  0.40s &  3.45s & 8.57 \\
    $^3$D4 \teamprint{3} & 22.54s & 51.90s & 2.30 & 1m 28s & 3m 22s & 2.28 \\
  $^4$D3 \teamprint{2,4} &  2.31s & 14.12s & 6.11 &  4.70s & 28.23s & 6.00 \\
$^6$D2 \teamprint{2,4,6} &  1.72s & 14.23s & 8.30 &  4.26s & 36.39s & 8.54 \\
  $^6$D2 \teamprint{4,6} &  3.80s & 19.99s & 5.26 & 11.09s &  1m 1s & 5.52 \\
    $^6$D2 \teamprint{6} & 30.20s & 1m 15s & 2.52 & 1m 11s &  3m 1s & 2.53 \\
\bottomrule
\end{tabular}}
    \caption{Comparison between runtime of our solver with and without TB-DAG postpcessing (\Cref{se:optimizations}).}
    \label{tab:postproc runtime effect}
\end{table}

\subsection{Postprocessing Time}
\Cref{tab:postproc times} shows the time required for the postprocessing. The postprocessing time is negligible in practice.

\begin{table}[H]
\centering
\colorlet{teamcol}{gray}
\newcommand{\cbox}[2]{\fcolorbox{white}[rgb]{1,1,1}{\phantom{28m 13s}\llap{#2}}}
\centering
\newcommand{\mymidrulegray}{\arrayrulecolor{gray}\mymidrule}
\newcommand{\mymidrule}{
\cmidrule(lr){1-5}
\cmidrule(lr){6-7}
\cmidrule(lr){8-9}
\cmidrule(lr){10-10}
\cmidrule(lr){11-15}
\arrayrulecolor{black}}
\newcommand{\unk}{\textcolor{black!30}{---}}
\setlength{\tabcolsep}{2mm}
\newcommand{\teamprint}[1]{\{\textcolor{teamcol}{#1}\}}
\begin{minipage}[t]{5cm}\begin{tabular}[t]{lr}
\toprule
      \multirow{2}{*}{\makebox[15mm][l]{{\bf Game} $\{\pmin\}$}} &
      \bf Postprocessing
      \\
      & \bf time
      \\
\midrule
    $^3$K3 \teamprint{3} & 0.00s \\
    $^3$K4 \teamprint{3} & 0.00s \\
    $^3$K6 \teamprint{3} & 0.00s \\
    $^3$K8 \teamprint{3} & 0.14s \\
  $^4$K5 \teamprint{3,4} & 0.00s \\
    $^4$K5 \teamprint{4} & 0.09s \\ %
    \bottomrule
\end{tabular}
\end{minipage}
\begin{minipage}[t]{5cm}\begin{tabular}[t]{lr}
\toprule
      \multirow{2}{*}{\makebox[15mm][l]{{\bf Game} $\{\pmin\}$}} &
      \bf Postprocessing
      \\
      & \bf time
      \\
\midrule
  $^3$L133 \teamprint{3} & 0.00s \\
  $^3$L143 \teamprint{3} & 0.01s \\
  $^3$L151 \teamprint{3} & 0.01s \\
  $^3$L153 \teamprint{3} & 0.10s \\
  $^3$L223 \teamprint{3} & 0.00s \\
  $^3$L523 \teamprint{3} & 1.13s \\ %
  \bottomrule
\end{tabular}
\end{minipage}
\begin{minipage}[t]{5cm}\begin{tabular}[t]{lr}
\toprule
      \multirow{2}{*}{\makebox[15mm][l]{{\bf Game} $\{\pmin\}$}} &
      \bf Postprocessing
      \\
      & \bf time
      \\
\midrule
$^4$L133 \teamprint{3,4} & 0.02s \\
    $^3$D3 \teamprint{3} & 0.03s \\
    $^3$D4 \teamprint{3} & 1.58s \\
  $^4$D3 \teamprint{2,4} & 0.56s \\
$^6$D2 \teamprint{2,4,6} & 0.40s \\
  $^6$D2 \teamprint{4,6} & 0.68s \\
    $^6$D2 \teamprint{6} & 1.45s \\
\bottomrule
\end{tabular}
\end{minipage}
    \caption{Time spent in the TB-DAG postprocessing operations (\Cref{se:optimizations}).}
    \label{tab:postproc times}
\end{table}

\subsection{Comparison between Our and Zhang and Sandholm's LP Size}

\Cref{tab:nnz comparison} shows a size comparison (in terms of number of nonzeros) between the TB-DAG and ZS22's tree decomposition, in terms of the number of nonzero entries in the resulting LP. The results show that the TB-DAG-based LP is between 2-5 times smaller (in terms of number of nonzeros in the program).

\begin{table}[H]
\centering
\colorlet{teamcol}{gray}
\newcommand{\cbox}[2]{\fcolorbox{white}[rgb]{1,1,1}{\phantom{28m 13s}\llap{#2}}}
\centering
\newcommand{\mymidrulegray}{\arrayrulecolor{gray}\mymidrule}
\newcommand{\mymidrule}{
\cmidrule(lr){1-5}
\cmidrule(lr){6-7}
\cmidrule(lr){8-9}
\cmidrule(lr){10-10}
\cmidrule(lr){11-15}
\arrayrulecolor{black}}
\newcommand{\unk}{\textcolor{black!30}{---}}
\setlength{\tabcolsep}{2mm}
\newcommand{\teamprint}[1]{\{\textcolor{teamcol}{#1}\}}
\begin{tabular}[t]{lrrr}
\toprule
      \multirow{2}{*}{\makebox[15mm][l]{{\bf Game} $\{\pmin\}$}} &
      \bf LP (Ours) &
      \bf ZS22 LP &
      \multirow{2}{*}{\bf Ratio}
      \\
      & size (nnz)
      & size (nnz)
      \\
\midrule
    $^3$K3 \teamprint{3} &     \num{1034} &     \num{2386} &  2.31 \\
    $^3$K4 \teamprint{3} &     \num{7073} &    \num{18810} &  2.66 \\
    $^3$K6 \teamprint{3} &   \num{338578} &  \num{1150838} &  3.40 \\
    $^3$K8 \teamprint{3} & \num{15569231} & \num{62574570} &  4.02 \\
  $^4$K5 \teamprint{3,4} &   \num{144252} &   \num{426297} &  2.96 \\
    $^4$K5 \teamprint{4} &  \num{4662152} & \num{21106658} &  4.53 \\
  $^3$L133 \teamprint{3} &    \num{56072} &   \num{126075} &  2.25 \\
  $^3$L143 \teamprint{3} &   \num{438893} &  \num{1195766} &  2.72 \\
  $^3$L151 \teamprint{3} &   \num{507718} &  \num{1425583} &  2.81 \\
\bottomrule
\end{tabular}~~
\begin{tabular}[t]{lrrr}
\toprule
      \multirow{2}{*}{\makebox[15mm][l]{{\bf Game} $\{\pmin\}$}} &
      \bf LP (Ours) &
      \bf ZS22 LP &
      \multirow{2}{*}{\bf Ratio}
      \\
      & size (nnz)
      & size (nnz)
      \\
\midrule
  $^3$L153 \teamprint{3} &  \num{3538848} & \num{11234573} &  3.17 \\
  $^3$L223 \teamprint{3} &    \num{56913} &   \num{112305} &  1.97 \\
  $^3$L523 \teamprint{3} &  \num{5161867} & \num{10507398} &  2.04 \\
$^4$L133 \teamprint{3,4} &   \num{388517} &   \num{785032} &  2.02 \\
    $^3$D3 \teamprint{3} &   \num{226526} &   \num{500665} &  2.21 \\
    $^3$D4 \teamprint{3} & \num{13940182} & \num{32755273} &  2.35 \\
  $^4$D3 \teamprint{2,4} &  \num{2699717} &  \num{5275196} &  1.95 \\
$^6$D2 \teamprint{2,4,6} &  \num{1076626} &  \num{1859959} &  1.73 \\
  $^6$D2 \teamprint{4,6} &  \num{2235785} &  \num{4749031} &  2.12 \\
    $^6$D2 \teamprint{6} &  \num{7501203} & \num{17635669} &  2.35 \\
\bottomrule
\end{tabular}
    \caption{size comparison (in terms of number of nonzeros) between the TB-DAG and ZS22's tree decomposition, in terms of the number of nonzero entries in the resulting LP}
    \label{tab:nnz comparison}
\end{table}

\section{CFR Convergence Plots}

In this section, we show the performance of each of the three CFR variants that we implemented to perform no-regret learning on the team belief DAG. As a rule of thumb, the predictive algorithm PCFR+ \citep{Farina21:Faster} is fastest when high precision (low team exploitability) is necessary. For low precision, DCFR \citep{Brown19:Solving} is often the fastest algorithm in practice, especially in certain variants of Kuhn poker.

\newcommand{\cfrone}[1]{\includegraphics[scale=.7]{cfr/#1.pdf}}
\newcommand{\cfr}[3]{%
    \noindent
    \begin{tikzpicture}%
        \node[anchor=west] at (12, 0) {\cfrone{#3}};
        \node[anchor=west] at (6, 0) {\cfrone{#2}};
        \node[anchor=west] at (0, 0) {\cfrone{#1}};
    \end{tikzpicture}%
}

\cfr{K34_110_CFR}{K36_110_CFR}{K38_110_CFR}

\cfr{K45_1100_CFR}{K45_1110_CFR}{L3133_110_CFR}

\cfr{L3143_110_CFR}{L3151_110_CFR}{L3153_110_CFR}

\cfr{L3223_110_CFR}{L3523_110_CFR}{L4133_1100_CFR}

\cfr{D33_110_CFR}{D34_110_CFR}{D43_1010_CFR}

\cfr{D62_101010_CFR}{D62_111010_CFR}{D62_111110_CFR}

\end{document}